\documentclass[aps,prc,groupedaddress,showpacs,manuscript]{revtex4-1}
\usepackage{graphicx}

\usepackage{colordvi}

\begin{document}

\title{The effect of internal magnetic field on collective flow in heavy ion collisions at intermediate energies}

\author {Yuliang Sun$\, ^{1}$\footnote{E-mail address: sunyl@zjhu.edu.cn},
Yongjia Wang$\, ^{1}$\footnote{E-mail address: wangyongjia@zjhu.edu.cn},
Qingfeng Li$\, ^{1,2}$\footnote{E-mail address: liqf@zjhu.edu.cn},
and
Fuqiang Wang$\, ^{1,3}$\footnote{E-mail address: fqwang@zjhu.edu.cn}}

\affiliation{
1) School of Science, Huzhou University, Huzhou 313000, P.R. China \\
2) Institute of Modern Physics, Chinese Academy of Sciences, Lanzhou 730000, P.R. China \\
3) Department of Physics and Astronomy, Purdue University, West Lafayette 47907, USA \\
\\
 }
\date{\today}

\begin{abstract}
The properties of nuclear matter under extreme conditions of high temperature, density and isospin-asymmetry have attracted wide attentions in recent years. At present, heavy ion reactions in combination with corresponding model simulations are one of the most important ways to investigate this subject. It is known that a strong magnetic field can be created in heavy ion collisions. However, its effect on the motion of charged particles is usually neglected in previous transport model simulations. In this work, within the Ultra-relativistic Quantum Molecular Dynamics (UrQMD) model, the temporal evolution and spatial distribution of the internal magnetic field are calculated. The magnetic field strength is found to reach about $eB\approx470$ MeV$^{2}$ ($B\approx8\times10^{16}$ G) for Au+Au collisions at $E_{\text{lab}}$=1 GeV/nucleon with impact parameter of 7 fm. The magnetic field in Cu+Au collisions exhibits somewhat different spatial distribution from that in Au+Au collisions. The magnetic field is found to affect the directed flow of pions at forward and backward rapidities to some extent, dependent of the impact parameter and beam energy while the effect on the elliptic flow is small. This suggests that, because $\pi$ mesons produced in heavy ion collisions at intermediate energies are considered as a sensitive probe for the nuclear symmetry energy, it is necessary to consider the effect of the internal magnetic field.
\end{abstract}



\maketitle
\section{Introduction}
The nuclear equation of state (EOS), a relationship involving the binding energy, nuclear matter density, isospin asymmetry and temperature, has been a continuous focus of research in nuclear physics~\cite{Blaizot:1980tw,Stoecker:1986ci,Baran:2004ih,Li:2008gp}. In particular, the isospin degree of freedom in the nuclear EOS has attracted wide attentions in recent decades. Although the EOS of isospin symmetric nuclear matter is relatively well constrained to a narrow region (its incompressibility $K_{0}$ is in the range of 200-260 MeV)~\cite{Danielewicz:2002pu,Dutra:2012mb,Fevre:2015fza,Wang:2018hsw}, the EOS of isospin asymmetric nuclear matter is still poorly understood~\cite{Tsang:2012se,Lattimer:2012xj,Li:2014oda,Russotto:2016ucm}. The largest uncertainty comes from the density dependence of the nuclear symmetry energy. Knowledge of the nuclear symmetry energy is important for understanding the properties of exotic nuclei, heavy-ion reactions with radioactive beams, and the structure of neutron stars~\cite{Baran:2004ih,Li:2008gp,Steiner:2004fi,Lattimer:2006xb,DiToro:2010ku,Lu:2016htm}. Heavy-ion collision experiments combined with model simulations are among the most important ways to investigate the EOS and the nuclear symmetry energy at high densities. A number of theoretical and experimental studies of the nuclear symmetry energy have been carried out. Until very recently, the nuclear symmetry energy at subnormal densities is relatively well constrained, but its high-density behavior is still poorly known~\cite{Lattimer:2012xj,Tsang:2012se,Xiao:2008vm,Russotto:2011hq,Cozma:2013sja,RocaMaza:2012mh,Zhang:2013wna,Brown:2013mga,Danielewicz:2013upa,Wang:2015kof}.

It was first pointed out by Rafelski and M\"{u}ller~\cite{PhysRevLett.36.517} that, in addition to strong electric field, strong magnetic field is also created in heavy-ion collisions (HICs). In sub-Coulomb barrier U+U collisions, the magnetic field was estimated to be on the order of $10^{14}$ G~\cite{PhysRevLett.36.517}. More recently, it was shown by Kharzeev et al.~\cite{KHARZEEV2008227} that HICs at the Relativistic Heavy Ion Collider (RHIC) and the Large Hadron Collider (LHC) can create the strongest magnetic field ever achieved in a terrestrial laboratory. For example, in noncentral Au + Au collisions at 100 GeV/nucleon, the maximal magnetic field can reach about $10^{18}$ G~\cite{KHARZEEV2008227}.

The strong magnetic field generated by heavy ion collisions in the relativistic energy region has recently attracted intense attention~\cite{KHARZEEV2008227,Deng:2012pc,Xu:2017zcn}, but its effect in the intermediate to low energy regions has not been thoroughly investigated. Ou and Li~\cite{Ou:2011fm} studied the temporal evolution and spatial distribution of internal electromagnetic fields in heavy-ion reactions within an isospin-dependent Boltzmann-Uhling-Uhlenbeck(IBUU) transport model, and found that the inner magnetic field had almost no effect on nucleon observables, but affected the pions at large rapidities. On the other hand, the two different frameworks of IBUU and Ultra-relativistic Quantum Molecular Dynamics (UrQMD) models have been simulated and compared in terms of sensitive probes of nuclear symmetry energy~\cite{GUO2013211,Guo:2014tua}. It was found that the neutron to proton ratio, the $ \pi^{-}/\pi^{+} $ ratio, and the isospin-sensitive transverse and elliptic flows from the two transport models are not always the same~\cite{GUO2013211,Guo:2014tua}. This motivated us to investigate the effects of the inner magnetic field in the UrQMD model. We study both the symmetric Au+Au and asymmetric Cu+Au collisions because the magnetic fields are expected to differ between these two systems. It is expected that anisotropic flow would be affected by the magnetic field, so we investigate the effects of the magnetic fields on directed flow and elliptic flow in Au+Au and Cu+Au collisions. This is particularly relevant for directed flow which has been extensively studied in Au+Au and Cu+Au at low energies~\cite{Voronyuk:2014rna,Toneev:2016bri}.

The rest of the paper is organized as follows. In the next section, we describe the UrQMD model and how the calculation of the internal magnetic field is implemented in the UrQMD model. The characteristics of the magnetic field and its effects on the collective flow, observables in heavy-ion collisions at intermediate energies are discussed in Sec. III, IV and V, respectively. Sec. VI summarizes our work.

\section{Magnetic field calculations in UrQMD}
The UrQMD model~\cite{Bass:1998ca,Bleicher:1999xi,Li:2011zzp,Li:2012ta} has been widely and
successfully used in the studies of \emph{pp}, \emph{p}A, and AA collisions over a large
range of energy from Bevalac and SIS up to the AGS, SPS, RHIC, and LHC.
At lower energies, the UrQMD model is based on principles analogous
to the quantum molecular dynamics (QMD) model \cite{aichelin91} in which each
nucleon is represented by
a Gaussian wave packet in phase space. The centroids ${\bf r}_i$ and ${\bf p}_i$
of a hadron $i$ in the coordinate and momentum spaces are
propagated according to Hamilton's equations of motion:

\begin{equation}\label{equ1}
{\bf \dot{r}}_i=\frac{\partial H}{\partial {\bf p}_i},
\hspace{1.5cm} \mathrm{and} \hspace{1.5cm} {\bf
\dot{p}}_i=-\frac{\partial H}{\partial {\bf r}_i}. \label{Hemrp}
\end{equation}
The total Hamiltonian $H$ consists of the kinetic energy $T$
and the effective two-body interaction potential energy $U$,
\begin{equation}
H=T+U.
\end{equation}

\begin{figure}[htbp]
\centering
\includegraphics[angle=0,width=0.7\textwidth]{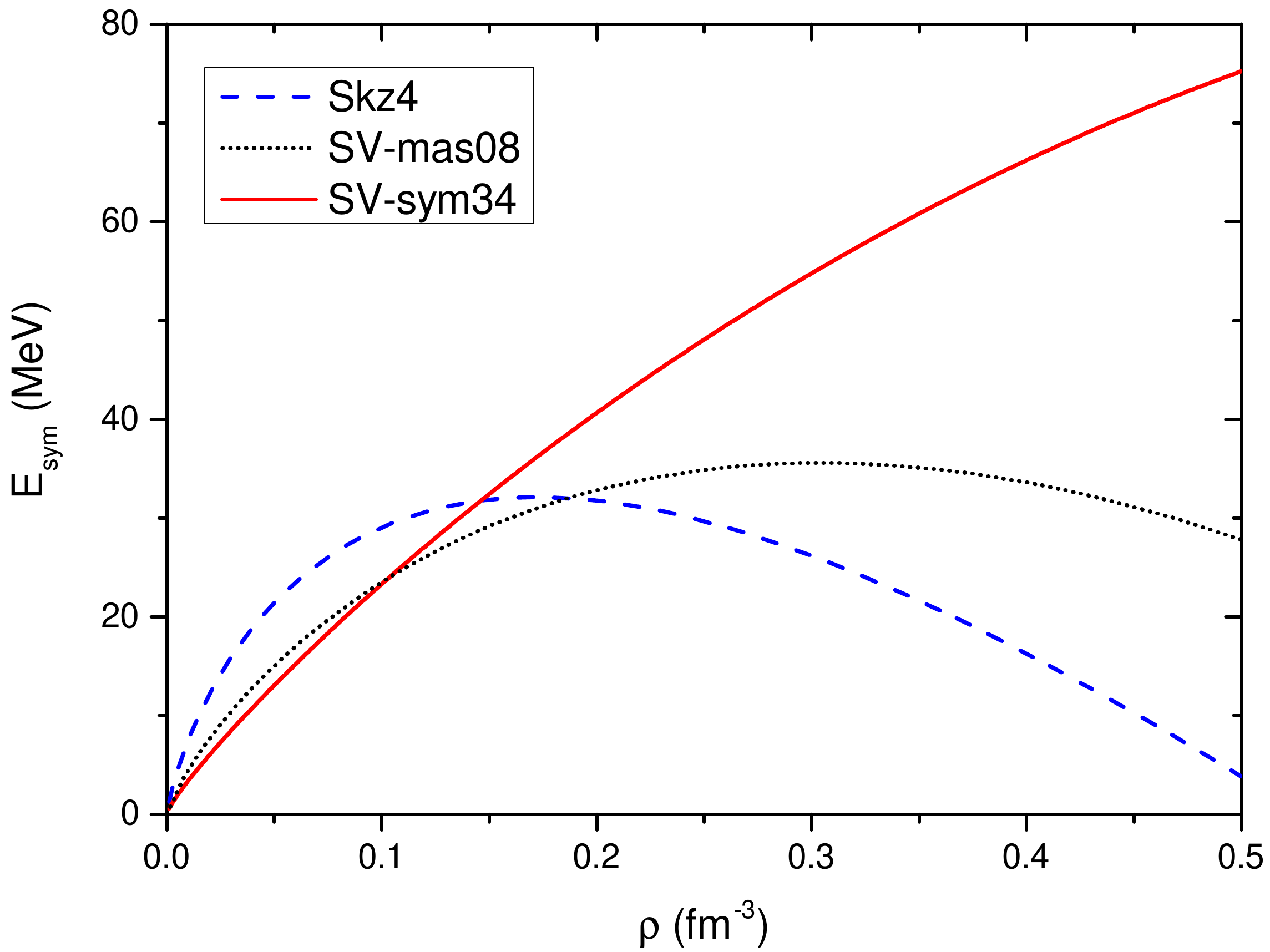}
\caption{\label{esym} (Color online) The nuclear symmetry energy obtained from Skz4, SV-mas08, and SV-sym34 are shown as a function of density.
}
\end{figure}

In this work, the mean field potential part is derived from the Skyrme energy density functional, and the SV-mas08~\cite{Dutra:2012mb} interaction with a corresponding incompressibility $K_{0}$=233 MeV is employed throughout the paper without explicit mention. In order to assess the impact of the magnetic filed on the extraction of the nuclear symmetry energy from isospin sensitive observables, two more interactions (Skz4 and SV-sym34) which give different values of the slope parameter of the nuclear symmetry energy are introduced. The nuclear symmetry energy obtained from those three interactions are displayed in Fig.\ref{esym}. The implementation of medium modifications for nucleon-nucleon elastic cross sections and Pauli blocking is consistent with Ref.~\cite{yongjia2014prc}. The experimental data of heavy-ion collisions in intermediate energies can be reproduced reasonably well by this model~\cite{yongjia2014prc,Wang:2016yti,Li:2016mqd,Zou:2016lpk,Liu:2018xvd,Du:2018ruo,Li:2018wpv}.

The coulomb field has already been considered in most transport models. The magnetic field at field point $\mathbf{r}_{i}$ (the $i$th particle's position) produced by particle motions can be calculated by the Li\'{e}nard-Wiechert potentials as in Ref.~\cite{Jackson}:
\begin{equation}\label{mag1}
e\mathbf{B}(\mathbf{r}_{i},t)=\frac{e^{2}}{4\pi\varepsilon_{0}c}\sum_{j=1,n}Z_{j}\frac{c^{2}-|\mathbf{v}_{j}|^{2}}{(c|\mathbf{R}_{j}|-\mathbf{R}_{j}\cdot \mathbf{v}_{j})^{3}}\mathbf{v}_{j}\times\mathbf{R}_{j}\bigg|_{\mathrm{Retarded}}.
\end{equation}
Here $Z_{j}$ is the charge number of the $j$th particle; $\mathbf{R}_{j}=\mathbf{r}_{i}-\mathbf{r}_{j}$ is the field point $\mathbf{r}_{i}$ relative to the position $\mathbf{r}_{j}$ of the particle $j$ moving at velocity $\mathbf{v}_{j}$, taken at retarded time~\cite{Jackson}. For constant $\mathbf{v}_{j}$ (e.g. of spectator protons), Eq.~(\ref{mag1}) reduces to~\cite{Jackson}
\begin{equation}\label{rel}
e\mathbf{B}(\mathbf{r}_{i},t)=\frac{e^{2}}{4\pi\varepsilon_{0}c}\sum_{j=1,n}Z_{j}\frac{c^{2}-|\mathbf{v}_{j}|^{2}}{((c|\mathbf{R}_{j}|)^{2}-(\mathbf{R}_{j}\times \mathbf{v}_{j})^{2})^{3/2}}\mathbf{v}_{j}\times\mathbf{R}_{j},
\end{equation}
where $\mathbf{r}_{j}$ and $\mathbf{v}_{j}$ are now taken at the time instant $t$.

\begin{figure}[htbp]
\centering
\includegraphics[angle=0,width=0.9\textwidth]{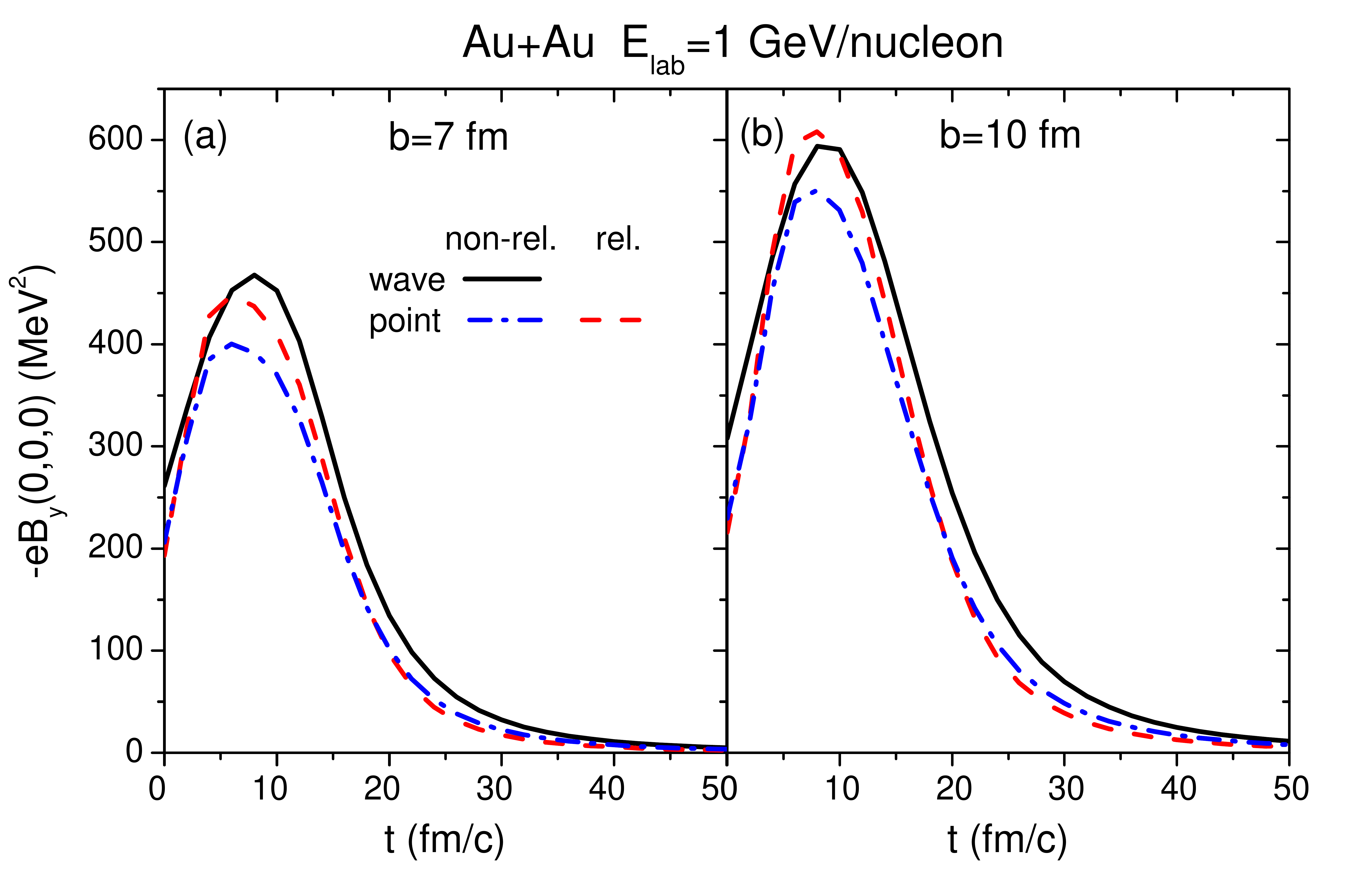}
\caption{\label{fig1} (Color online) Temporal evolution of the magnetic field component $-eB_{y}(0, 0, 0)$ produced by spectator protons in Au+Au collisions at the beam energy $E_{\text{lab}}$=1 GeV/nucleon with (a) $b=7$ fm and (b) $b=10$ fm. The legend \textquotedblleft non-rel. wave\textquotedblright, \textquotedblleft non-rel. point\textquotedblright~and \textquotedblleft rel. point\textquotedblright~indicate the results obtained using Eqs.~(\ref{mag2}),~(\ref{non-rel}) and~(\ref{rel}), respectively.
}
\end{figure}

For participant charged particles, the retardation effect is tedious to implement because of the changes of their velocities. One approach is to use the non-relativistic approximation by assuming $\mathbf{v}_{j}/c\ll1$ in Eq.~(\ref{mag1}), namely
\begin{equation}\label{non-rel}
e\mathbf{B}(\mathbf{r}_{i},t)=\frac{e^{2}}{4\pi\varepsilon_{0}c^{2}}\sum_{j=1,n}Z_{j}\frac{1}{|\mathbf{R}_{j}|^{3}}\mathbf{v}_{j}\times\mathbf{R}_{j}.
\end{equation}
In order to see how large the effect of relativity is, we compare the values of $-eB_{y}(0, 0, 0)$ produced by spectators protons in Au+Au collisions at the beam energy $E_{\text{lab}}$=1 GeV/nucleon calculated by Eq.~(\ref{rel}) and by the non-relativistic approximation of Eq.~(\ref{non-rel}). These are shown as the red dash curves and blue dash dot curves in Fig.~\ref{fig1}, respectively. The $-eB_{y}$ is plotted in unit of MeV$^{2}$, which is equal to $1.7\times10^{14}$ G. The effect of relativity is approximately $+20\%$ at time $t\lesssim20$ fm/c. At later times, the effect reverses sign. The relative effect is larger, but since the magnetic field is relatively small, the effect is not expected to affect our results significantly. Note, in UrQMD, $t=0$ is always defined to be the time instant when the surface distance between the two nuclei is 1.6 fm.

Since particles are treated as point-like, singularities appear when $\mathbf{R}_{j}=0$. In order to avoid the singularity by effectively considering the size of the particles, the magnetic field produced by particle $j$ is ignored in the above calculation when $|\mathbf{R}_{j}|<1$ fm (using smaller cutoff values yielded consistent results). To avoid the singularity problem all together, and arguably more physically motivated, one many use Gaussian wave packets instead of point-like particles. To consider wave packet in calculating the magnetic field by particle $j$ at field point $\mathbf{r}_{i}$, we first obtain the vector potential~\cite{PhysRevC.83.054911}:
\begin{equation}\label{magA}
\mathbf{A}_{j}(\mathbf{r}_{i},t)=\frac{e}{4\pi \varepsilon_{0} c^{2}}\mathbf{v}_{j}\int\rho_{i}(\mathbf{r},t)\frac{1}{\mathbf{r}-\mathbf{r'}}\rho_{j}(\mathbf{r'},t)d\mathbf{r}d\mathbf{r'}.
\end{equation}
Here
\begin{equation}
\rho_{i}(\mathbf{r},t)=\frac{1}{(2\pi \sigma_{r}^2)^{3/2}}\exp{\left (-\frac{(\mathbf{r}-\mathbf{r}_{i}(t))^{2}}{2\sigma_{r}^2} \right )},
\end{equation}
with $\sigma_{r}^2=2$ fm$^2$ being the width parameter of Gaussian. The magnetic field is then given by
\begin{equation}\label{magB}
\mathbf{B}_{j}(\mathbf{r}_{i},t)=\bigtriangledown\times\mathbf{A}_{j}(\mathbf{r}_{i},t),
\end{equation}
and the total magnetic field
\begin{eqnarray*}
\mathbf{B}(\mathbf{r}_{i},t)=\mathop{\sum}_{j}\mathbf{B}_{j}(\mathbf{r}_{i},t)
\end{eqnarray*}
produced by all particles can be obtained. Because UrQMD has already computed the electric potential in its coding implementation,
\begin{equation}
e\Phi_{j}(\mathbf{r}_{i},t)=\frac{e^2}{4\pi\varepsilon_{0}}\times\int\rho_{i}(\mathbf{r},t)\frac{1}{\mathbf{r}-\mathbf{r'}}\rho_{j}(\mathbf{r'},t)d\mathbf{r}d\mathbf{r'},
\end{equation}
the magnetic field strength can be readily calculated by
\begin{equation}\label{mag2}
e\mathbf{B}_{j}(\mathbf{r}_{i},t)=\bigtriangledown\times e\mathbf{A}_{j}(\mathbf{r}_{i},t)=\bigtriangledown\times(\frac{e}{c^2}\Phi_{j}(\mathbf{r}_{i},t)\mathbf{v}_{j})\\
=-\frac{e}{c^2}\mathbf{v}_{j}\times\bigtriangledown\Phi_{j}(\mathbf{r}_{i},t).
\end{equation}
We note here that the relativistic effect is not considered in this method. $-eB_{y}(0, 0, 0)$ produced by spectators protons in Au+Au collisions at the beam energy $E_{\text{lab}}$=1 GeV/nucleon calculated by Eq.~(\ref{mag2}) are shown as the black solid curves in Fig.~\ref{fig1}.

\begin{figure}[htbp]
\centering
\includegraphics[angle=0,width=0.9\textwidth]{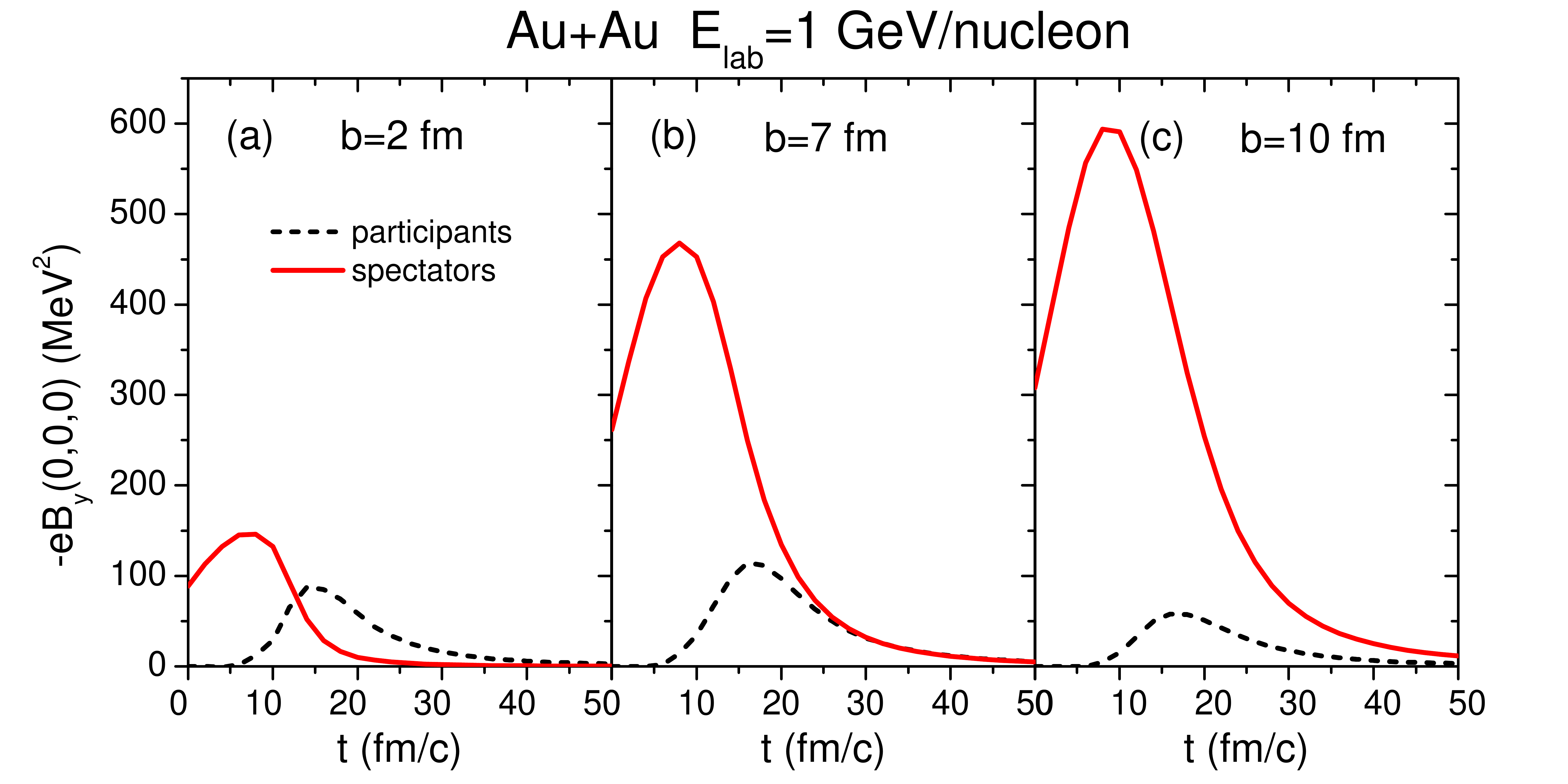}
\caption{\label{fig2} (Color online) The magnetic field strengths $-eB_{y}(0, 0, 0)$ produced by participants or spectators in Au+Au collisions at the beam energy $E_{\text{lab}}$=1 GeV/nucleon with impact parameter $b$=2, 7, 10 fm.
}
\end{figure}

In principle, the magnetic field strength should be calculated by the relativistic formula of Eq.~(\ref{mag1}). However, it is hard to implement for participant charged particles because their velocities change over time. We thus examine next the relative contributions to the magnetic field from participants and spectators using the non-relativistic approx imation of Eq.~(\ref{mag2}). Figure~\ref{fig2} is the magnetic field strengths $-eB_{y}(0, 0, 0)$ produced by participants and spectators, respectively, in Au+Au collisions at the beam energy $E_{\text{lab}}$=1 GeV/nucleon. The magnetic field strengths produced by participants and spectators are both dependent of the impact parameter ($b$) because the numbers of participants and spectators vary with $b$. The magnetic field strength produced by participants is considerable, especially in central collision, and thus can not be neglected. We therefore opt for the non-relativistic wave-packet calculations Eq.~(\ref{mag2}) in this work, with the understanding that the relativistic correction is on the order of $+20\%$ at $t\lesssim15$ fm/c when the magnetic field is appreciable. Since, as we will show, the influence of magnetic field on our observables is generally small, less than 10\%, the approximation on the magnetic field introduces an uncertainty on the results of this work only by a few percent, without changing the qualitative conclusions.

The Lorentz force on the charged particle $i$ can be written as
\begin{equation}\label{Lorentz}
\mathbf{F}_{i(mag)}=Z_{i}\mathbf{v}_{i}\times e\mathbf{B}(\mathbf{r}_{i},t).
\end{equation}
To take into account the effect of inner magnetic field in the UrQMD model, in the quasiparticle approximation, the Lorentz force is added to Eq.~(\ref{equ1}), namely
\begin{equation}\label{equ1}
{\bf \dot{r}}_i=\frac{\partial H}{\partial {\bf p}_i},
\hspace{1.5cm} \mathrm{and} \hspace{1.5cm} {\bf
\dot{p}}_i=-\frac{\partial H}{\partial {\bf r}_i}+\mathbf{F}_{i(mag)}.
\end{equation}

\section{Magnetic field results in heavy-ion collisions}
We take the $z$-axis as the beam direction and $x$-axis as the impact parameter direction. In the limit of smooth nuclear density, because of symmetry, only the $y$ component of the magnetic field is non-vanishing. On event-by-event basis, because of fluctuations, all $x, y, z$ components of the magnetic field are present~\cite{Deng:2012pc}. In our simulation, all three components of the magnetic field are calculated and included in the Lorentz force of Eq.~(\ref{Lorentz}). Averaged over many events, the $x, z$ components of the magnetic field vanish, and only the $y$ component remains. For the flow variables we study which are event-averaged quantities, only the $B_{y}$ component will have an effect. In this section we depict the $y$ component of the magnetic field averaged over many events.

\begin{figure}[htbp]
\centering
\includegraphics[angle=0,width=1.0\textwidth]{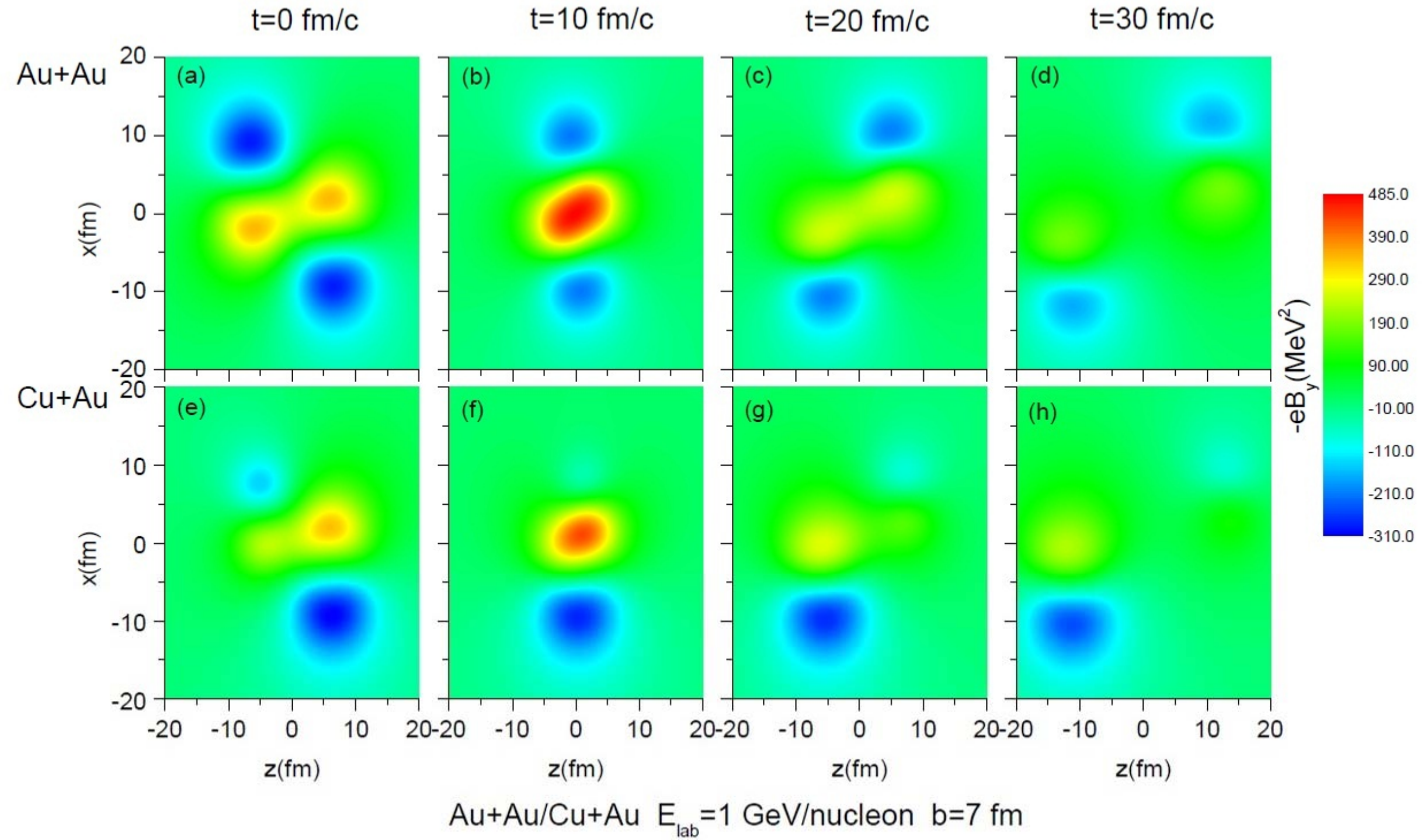}
\caption{\label{fig3} (Color online) Distributions of the magnetic field strength $-eB_{y}$ in the reaction plane at $t=0, 10, 20, 30$ fm/$c$ in Au+Au (top) and Cu+Au (bottom) collisions at the beam energy $E_{\text{lab}}$=1 GeV/nucleon with impact parameter $b=7$ fm.
}
\end{figure}

Figure~\ref{fig3} shows the magnetic field strength $-eB_{y}$ in the $y=0$ plane at $t=0, 10, 20, 30$ fm/$c$ in Au+Au (top) and Cu+Au (bottom) collisions at the beam energy $E_{\text{lab}}$=1 GeV/nucleon with impact parameter $b=7$ fm. In the overlap zone ($|x|\lesssim5$ fm), the magnetic fields generated by the two spectators add up as they are both in the $-y$ direction. The strength of the magnetic field peaks when the two nuclei reach the maximum compression and drops when the nuclei depart from each other. The magnetic fields in the outer regions ($|x|\gtrsim5$ fm) partially cancel each other as the magnetic fields from the two spectators are in the opposite directions in $y$. In Cu+Au collisions, the magnetic field generated by the Cu nucleus is smaller than that by the Au nucleus at the same distances of the field point, so the two magnetic fields largely cancel each other in the $+x$ direction but only partially at $-x$ direction. The largest magnitude magnetic field is not at the origin.

\begin{figure}[htbp]
\centering
\includegraphics[angle=0,width=0.9\textwidth]{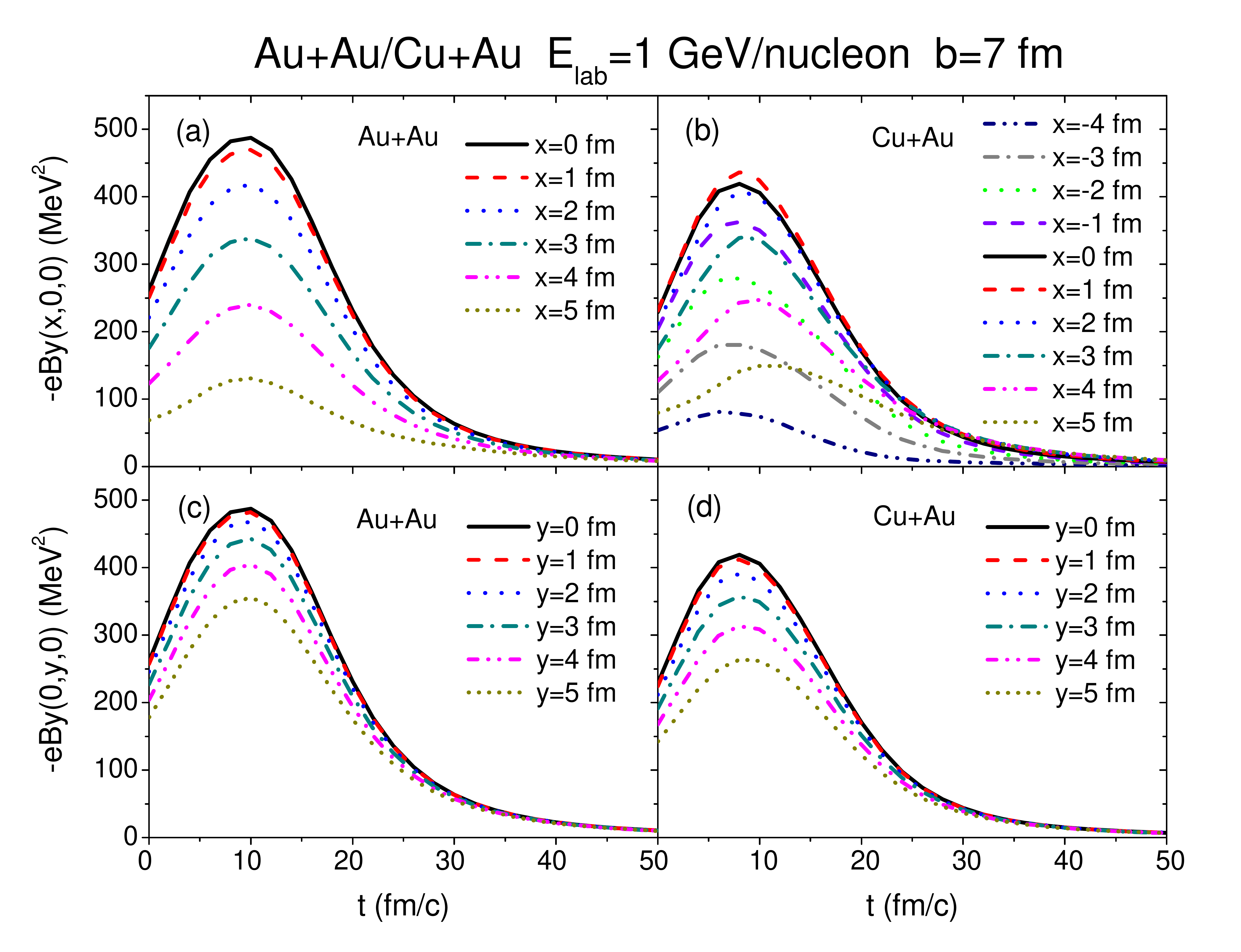}
\caption{\label{fig4} (Color online) Temporal evolution of the magnetic field $-eB_{y}$ at the point ($x$, 0, 0) in (a) Au+Au and (b) Cu+Au, and at the point (0, $y$, 0) in (c) Au+Au and (d) Cu+Au at the beam energy $E_{\text{lab}}$=1 GeV/nucleon with impact parameter $b=7$ fm.
}
\end{figure}

The magnetic field $-eB_{y}(x,0,0)$ is shown as a function of time for various transverse coordinate $x$ in Fig.~\ref{fig4}(a) Au+Au and (b) Cu+Au. The magnetic field strength increases with time, reaches about $eB\approx470$ MeV$^{2}$ ($B\approx8\times10^{16}$ G) at $t=10$ fm/$c$, and then decreases with increasing time. The magnetic field is strongest in the central region and decreases with increasing $x$. There are several differences in Cu+Au compared to Au+Au: the largest magnetic field strength is at $x\approx1$ fm instead of $x=0$; the peaks of the magnetic field strength appear later in the $x>0$ region (Cu side) and earlier in the $x<0$ region (Au side); and the magnetic field strengths at some places of $x>0$ in Cu+Au are stronger than those in Au+Au, although the magnetic field strength in the center is weaker than that in Au+Au.
The magnetic field $-eB_{y}(0,y,0)$ as a function of time for various $y$ positions is present in Fig.~\ref{fig4}(c) Au+Au and (d) Cu+Au. The magnetic field strength decreases with increasing $y$, the distance from the reaction plane.

\begin{figure}[htbp]
\centering
\includegraphics[angle=0,width=0.7\textwidth]{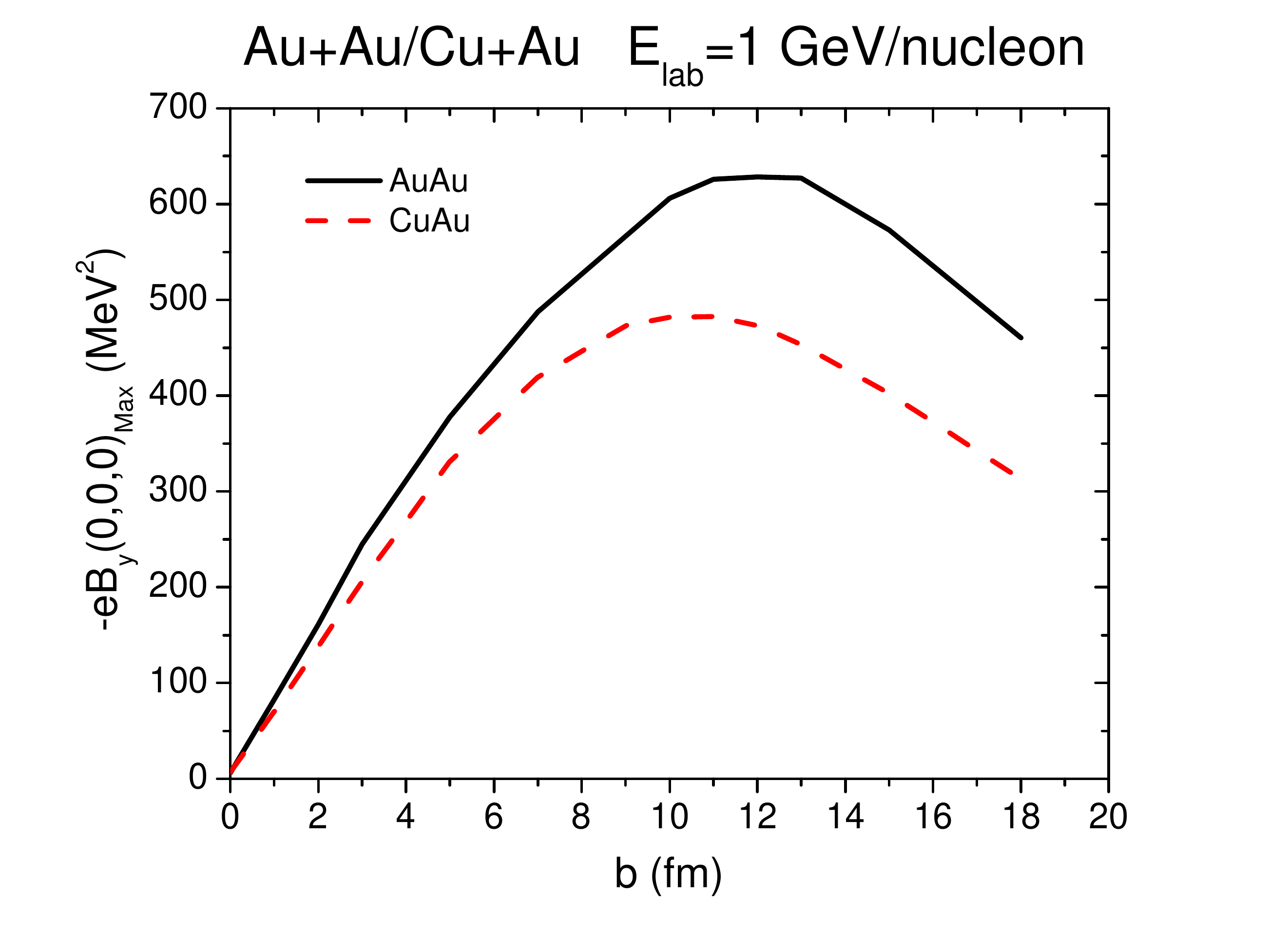}
\caption{\label{b-ByMax} (Color online) The peak value of the time evolution of $-eB_{y}(0, 0, 0)$ as a function of the impact parameter $b$ in Au+Au/Cu+Au collisions at the beam energy $E_{\text{lab}}$=1 GeV/nucleon.
}
\end{figure}

Shown in Fig.~\ref{b-ByMax} is the maximum of $-eB_{y}(0, 0, 0)$ as functions of the impact parameter $b$ in Au+Au/Cu+Au collisions at the beam energy $E_{\text{lab}}$=1 GeV/nucleon. The strength of the magnetic field increases with increasing impact parameter $b$, and then decreases with increasing $b$. The dependence is understood because of the interplay of the number of spectator protons and the distances of the center (0,0,0) from those spectator protons.

\begin{figure}[htbp]
\centering
\includegraphics[angle=0,width=0.8\textwidth]{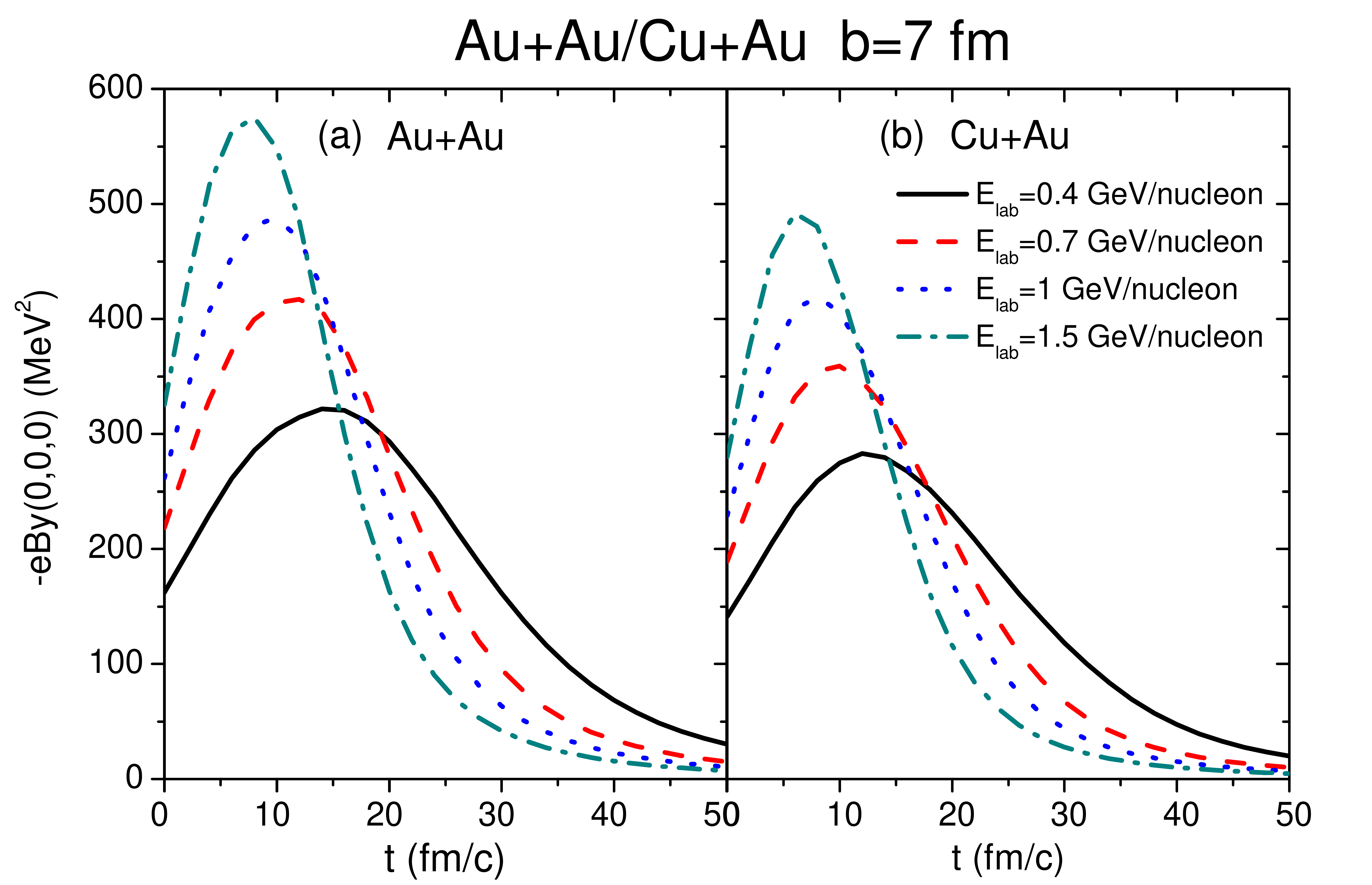}
\caption{\label{fig6} (Color online) Beam energy dependence of $-eB_{y}$(0, 0, 0) in (a) Au+Au and (b) Cu+Au collisions with $b=7$ fm.
}
\end{figure}

\begin{figure}[htbp]
\centering
\includegraphics[angle=0,width=0.7\textwidth]{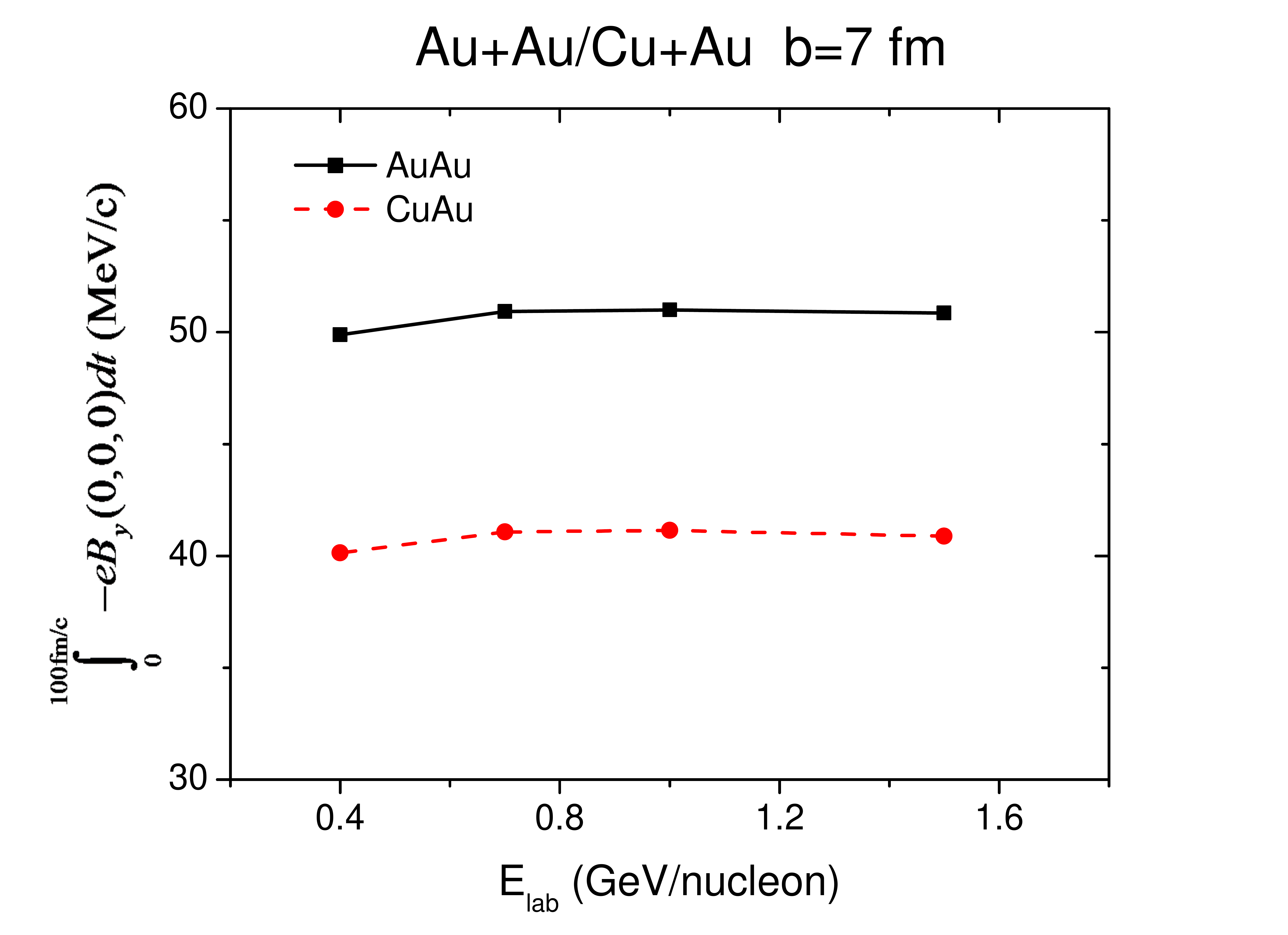}
\caption{\label{E-intB} (Color online) The integral of $\int^{\infty}_{0}-eB_{y}(0, 0, 0)dt\approx\int^{100\text{fm/c}}_{0}-eB_{y}(0, 0, 0)dt$ as a function of beam energy in Au+Au and Cu+Au collisions with $b=7$ fm.
}
\end{figure}

Figure~\ref{fig6} is the beam energy dependence of $-eB_{y}$(0, 0, 0) in Au+Au/Cu+Au collisions with $b=7$ fm. The magnetic field reaches maximum at shorter times at higher energies as expected. The magnetic field increases with increasing beam energy, while the duration of the magnetic field decreases because the spectators leave the collision region more quickly at higher beam energies. Both the magnetic field strength and duration are important for observable effects. We thus show in Fig.~\ref{E-intB} the more relevant integral quantity,
\begin{eqnarray*}
\int^{\infty}_{0}-eB_{y}(0, 0, 0)dt\approx\int^{100\text{fm/c}}_{0}-eB_{y}(0, 0, 0)dt,
\end{eqnarray*}
as a function of beam energy in Au+Au/Cu+Au collisions with $b=7$ fm. Only weak energy dependence is observed for this quantity.

\section{Magnetic effects on the collective flow of pions}

\begin{figure}[htbp]
\centering
\includegraphics[angle=0,width=0.6\textwidth]{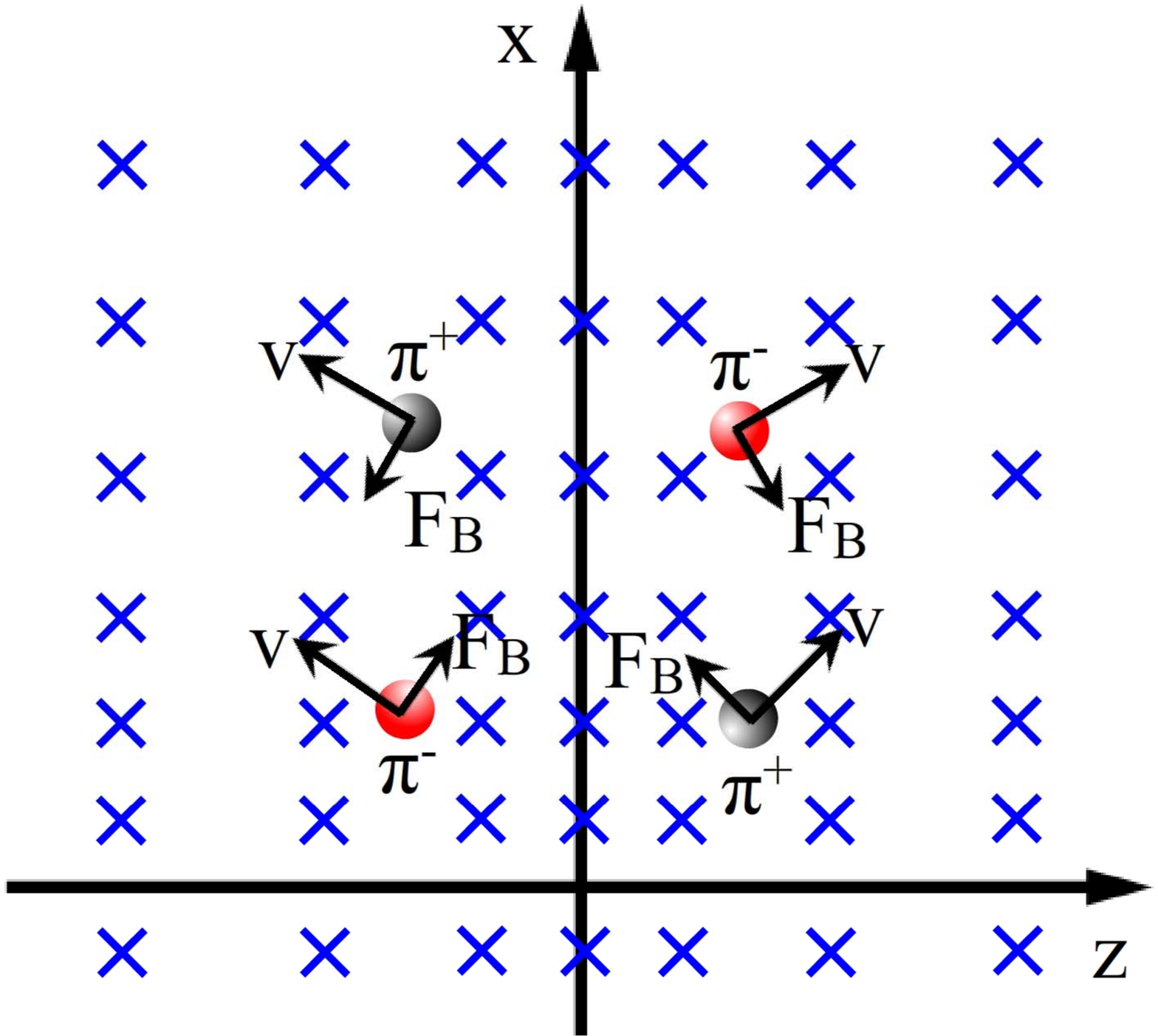}
\caption{\label{explain} (Color online) A sketch illustrating the Lorentz force on charged pions.
}
\end{figure}

\begin{figure}[htbp]
\centering
\includegraphics[angle=0,width=0.9\textwidth]{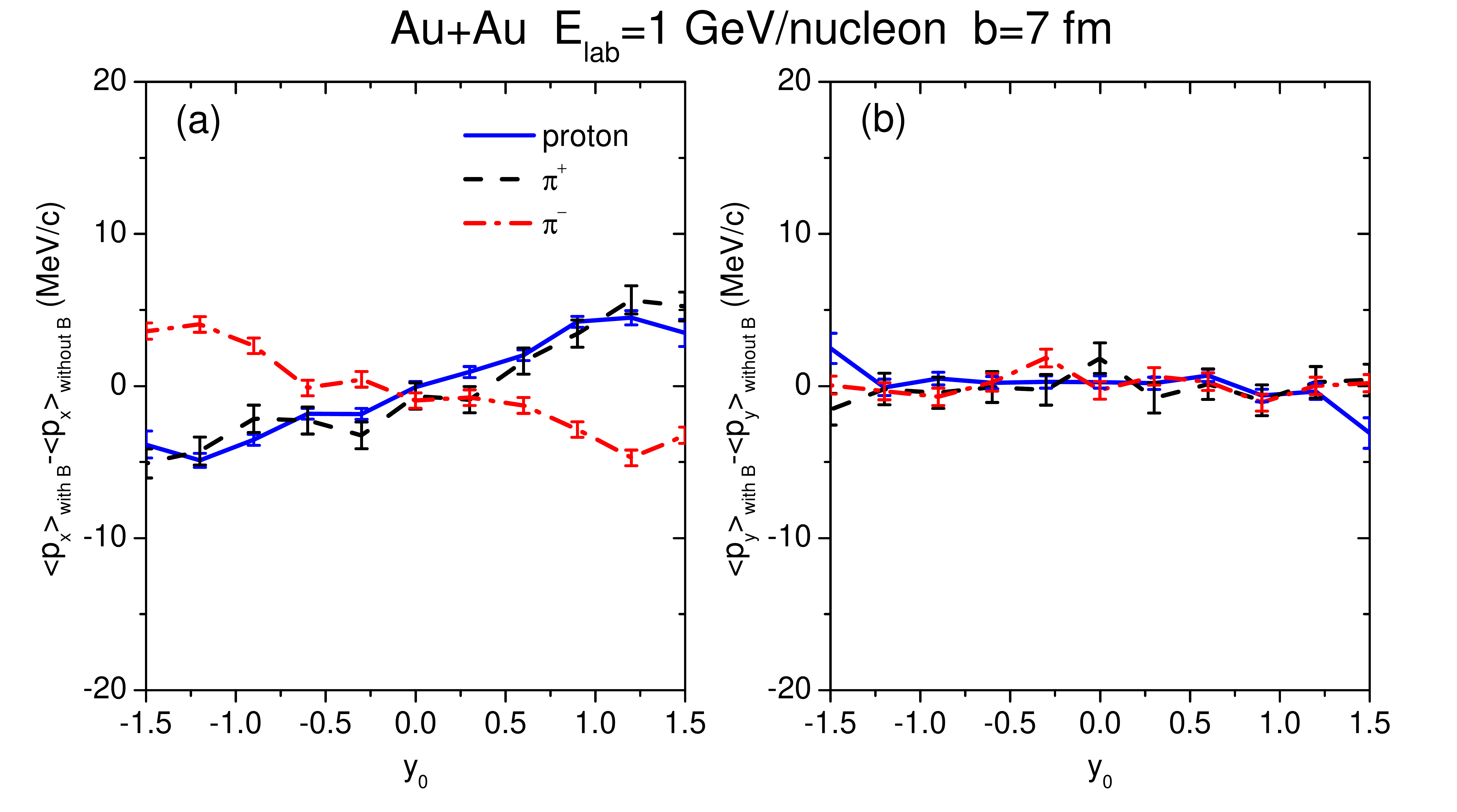}
\caption{\label{diff-pxpy} (Color online) The change of protons and pions' (a) $\langle p_{x}\rangle$ and (b) $\langle p_{y}\rangle$ by the Lorentz force as a function of normalized center-of-mass rapidity in Au+Au collision at $E_{\text{lab}}$=1 GeV/nucleon with $b=7$ fm.
}
\end{figure}

\begin{figure}[htbp]
\centering
\includegraphics[angle=0,width=0.9\textwidth]{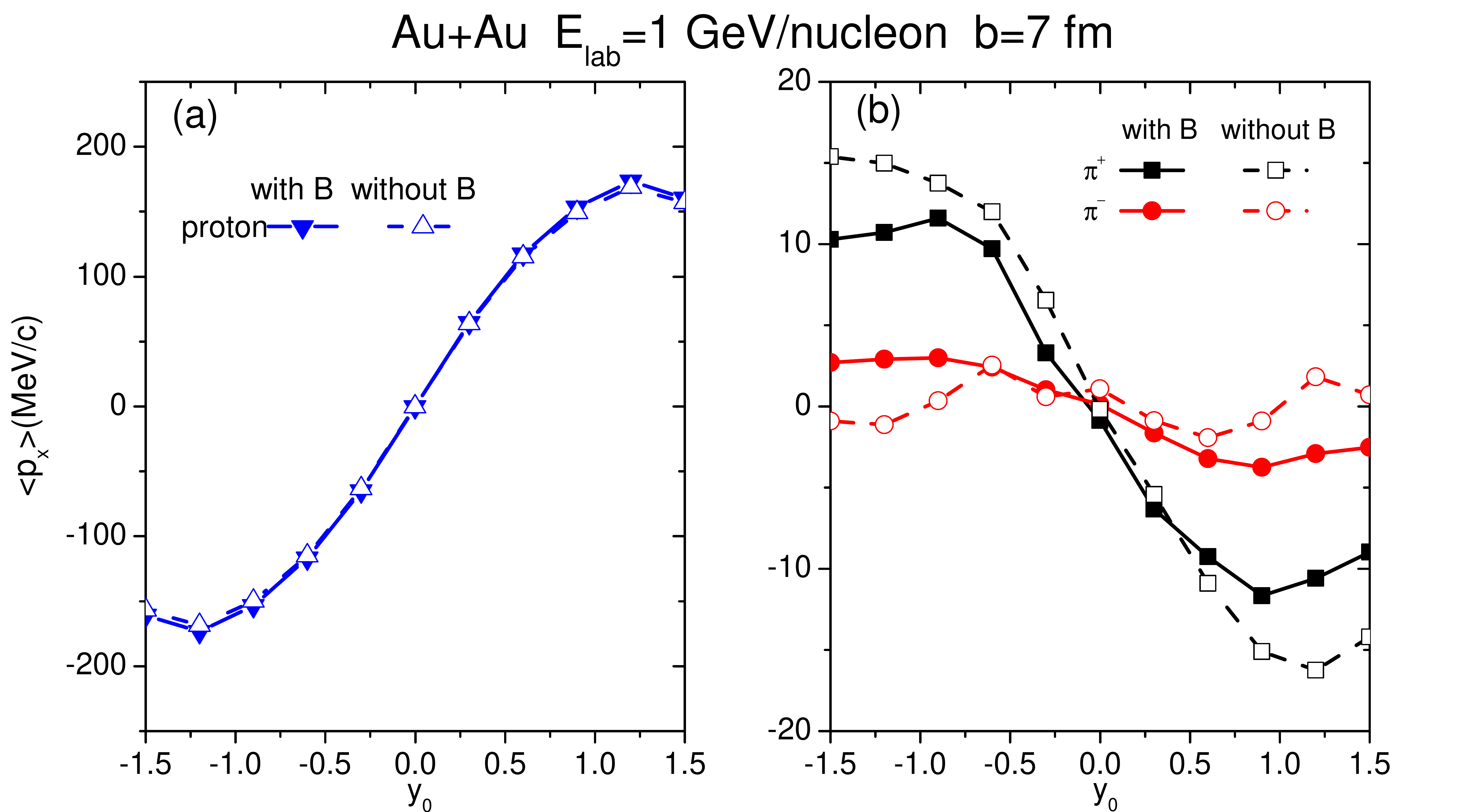}
\caption{\label{y0-pxpy} (Color online) $\langle p_{x}\rangle$ of (a) protons and (b) pions as a function of normalized center-of-mass rapidity in Au+Au collision at $E_{\text{lab}}$=1 GeV/nucleon with $b=7$ fm.
}
\end{figure}

From the expression of the Lorentz force of Eq.~(\ref{Lorentz}), it is easy to see that the main component of the Lorentz force is in the $x$ direction because the average magnetic filed is in the $y$ direction. The $x$ component of the average momentum, $\langle p_{x}\rangle$, is thus a quantity most likely to be affected. This is illustrated in Fig.~\ref{explain}. The Lorentz force deflects the $\pi^{+}$ direction of motion; $p_{x}$ increase (decrease) if $p_{z}>0$ ($p_{z}<0$). The effect on $\pi^{-}$ is opposite. Figure~\ref{diff-pxpy} (a) shows the change in $\langle p_{x}\rangle$ due to the magnetic field for protons and pions as a function of the center-of-mass rapidity $y=\frac{1}{2}\text{ln}(E+p_z)/(E-p_z)$ in Au+Au collision at $E_{\text{lab}}$=1 GeV/nucleon with $b=7$ fm. We use the scaled units $y_0\equiv y/y_{pro}$ as done in \cite{Reisdorf:2010aa}, and the subscript \emph{pro} denotes the incident projectile in the center-of-mass system.
The larger the rapidity (hence and average $p_{z}$), the larger the change in $\langle p_{x}\rangle$ as expected from the Lorentz force. The changes in forward and backward rapidities are on the order of 5 MeV/c. This is smaller than shown in Fig.~\ref{E-intB} because there the time integral is for the maximum magnetic field at the center of the collision zone.
The change for $\pi^{\pm}$ are opposite in sign because of the opposite charges. The changes for proton and $\pi^{+}$ are similar because the $z$ component of the velocity, $\frac{p_{z}}{E}=\text{tanh}~y$, is independent of particle species. Note that even though the $x$ and $z$ component of the magnetic field, due to fluctuations, do not vanish from point to point on an event-by-event basis, the averages of their effects diminish. This is shown by the zero change in $\langle p_{y}\rangle$, as example, in Fig.~\ref{diff-pxpy} (b).
Fig.~\ref{y0-pxpy} shows the $\langle p_{x}\rangle$ of (a) protons and (b) pions as a function of rapidity in Au+Au collision at $E_{\text{lab}}$=1 GeV/nucleon with $b=7$ fm. The change in the pion $p_{x}$ is obvious. The change in proton $\langle p_{x}\rangle$ is not noticeable because the proton $\langle p_{x}\rangle$ is 20 times than pion (note the different coordinate scales in Fig.~\ref{y0-pxpy}). The proton $\langle p_{x}\rangle$ is large because of its large mass and the large nuclear force they experience, much larger than the magnetic Lorentz force.
In present model, the motion pions only affected through pion-hadron collisions and the Coulomb field. Therefore, the effect of magnetic field on the pion flow is more obvious than that on the flow of proton.

\begin{figure}[htbp]
\centering
\includegraphics[angle=0,width=0.9\textwidth]{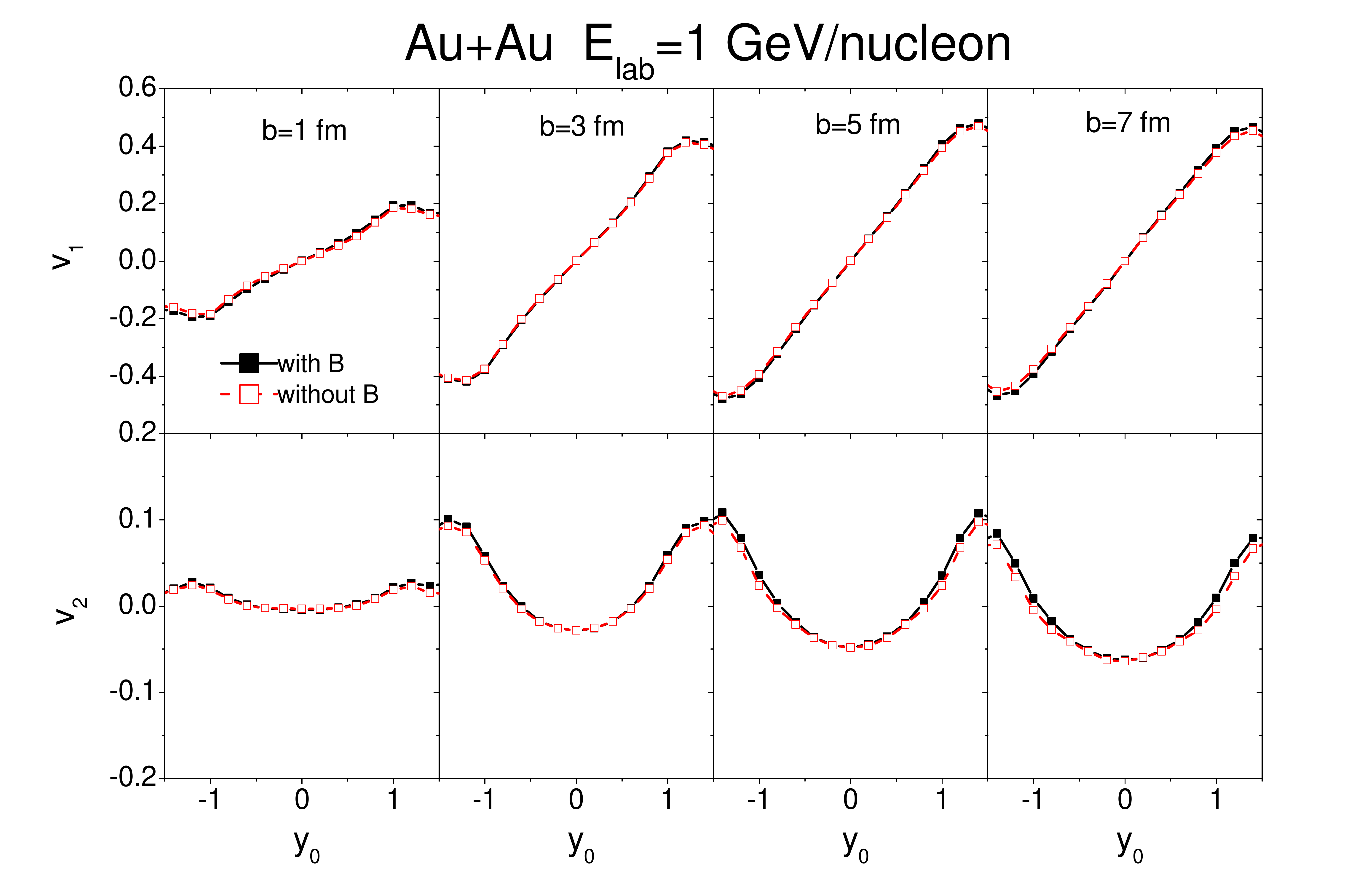}
\caption{\label{AuAu-p-v1v2} (Color online) $v_{1}$ and $v_{2}$ of protons as a function of normalized center-of-mass rapidity in Au+Au collisions at $E_{\text{lab}}$=1 GeV/nucleon with impact parameters of 1, 3, 5, 7 fm.
}
\end{figure}

Since the $\langle p_{x}\rangle$ is affected by the magnetic field, we examine the influence of the inner magnetic field on the directed flow ($v_{1}=\langle \frac{p_{x}}{\sqrt{p^{2}_{x}+p^{2}_{y}}}\rangle$) and the elliptic flow ($v_{2}=\langle \frac{p^{2}_{x}-p^{2}_{y}}{p^{2}_{x}+p^{2}_{y}}\rangle$). Figure~\ref{AuAu-p-v1v2} shows $v_{1}$ and $v_{2}$ of protons as functions of normalized center-of-mass rapidity in Au+Au collisions at $E_{\text{lab}}$=1 GeV/nucleon with impact parameters of 1, 3, 5, 7 fm. No much magnetic effect is observed on protons which is consistent with Fig.~\ref{y0-pxpy} (a).

\begin{figure}[htbp]
\centering
\includegraphics[angle=0,width=0.9\textwidth]{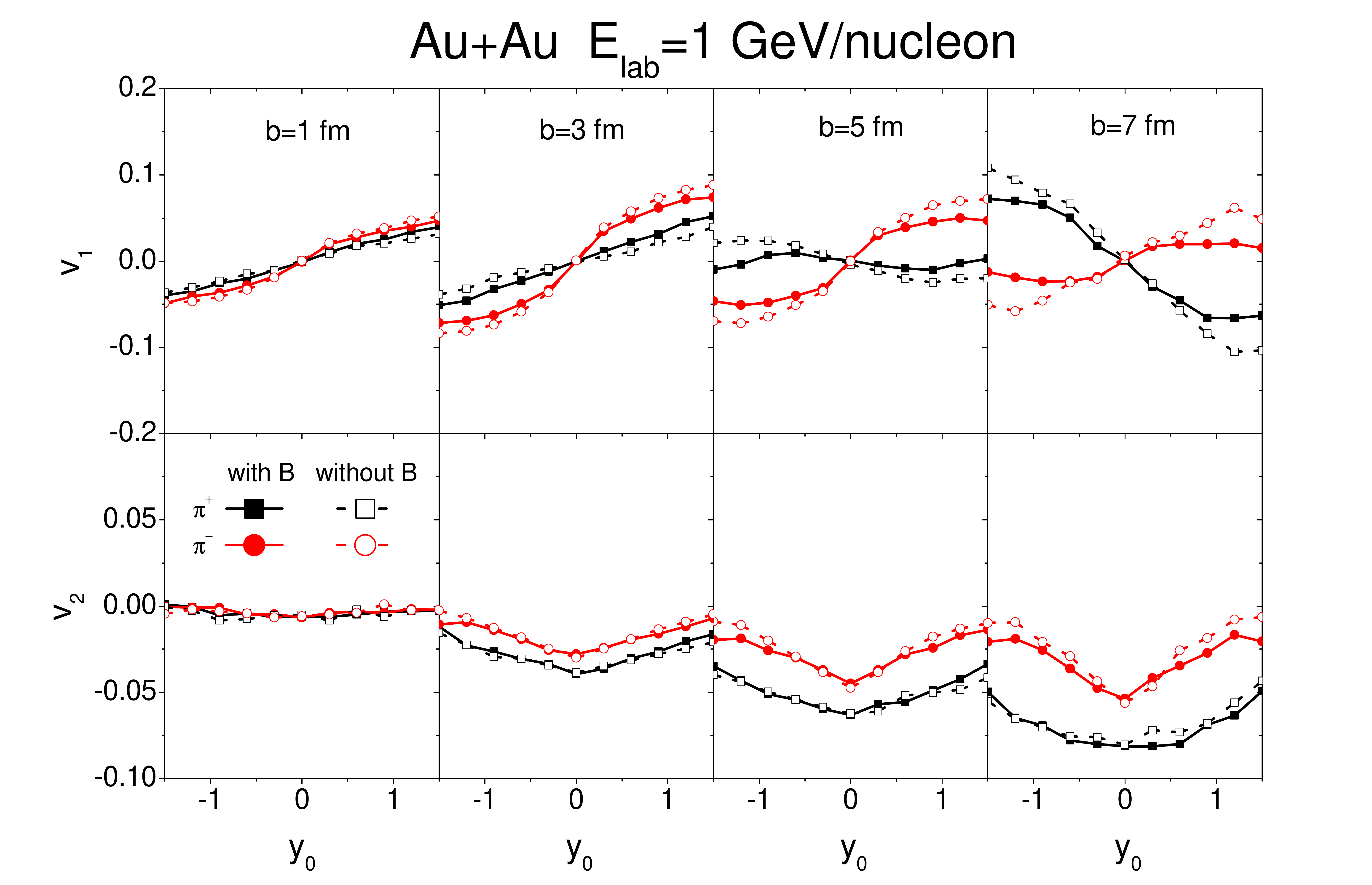}
\caption{\label{AuAub-pi-v1v2} (Color online) $v_{1}$ and $v_{2}$ of charged pions as a function of normalized center-of-mass rapidity in Au+Au collisions at $E_{\text{lab}}$=1 GeV/nucleon with impact parameters of 1, 3, 5, 7 fm.
}
\end{figure}

\begin{figure}[htbp]
\centering
\includegraphics[angle=0,width=0.9\textwidth]{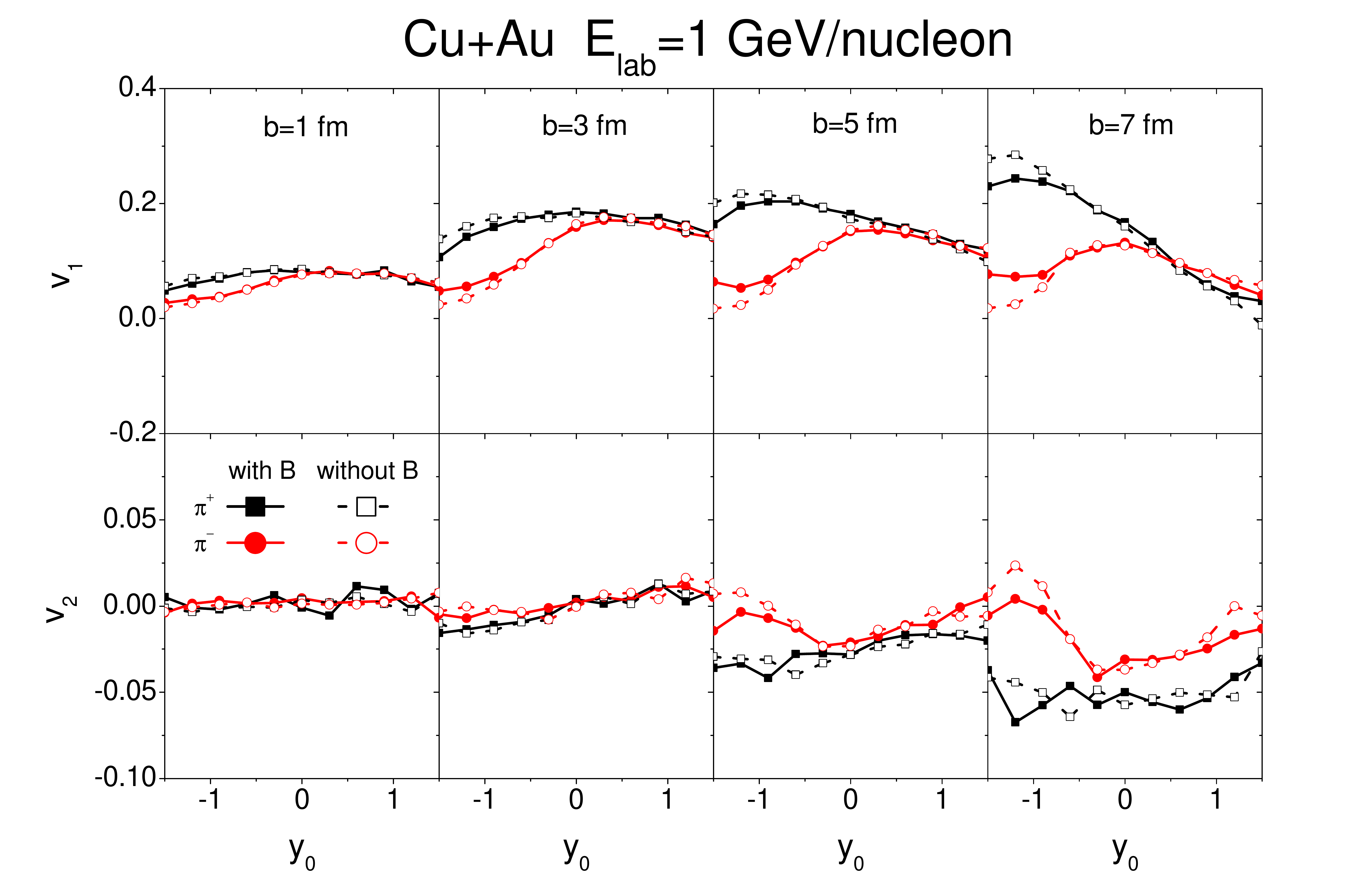}
\caption{\label{CuAub-pi-v1v2} (Color online) $v_{1}$ and $v_{2}$ of charged pions as a function of normalized center-of-mass rapidity in Cu+Au collisions at $E_{\text{lab}}$=1 GeV/nucleon with impact parameters of 1, 3, 5, 7 fm.
}
\end{figure}

Figure~\ref{AuAub-pi-v1v2},~\ref{CuAub-pi-v1v2} show $v_{1}$ and $v_{2}$ of charged pions as a function of normalized center-of-mass rapidity in Au+Au and Cu+Au collisions at $E_{\text{lab}}$=1 GeV/nucleon with impact parameters of 1, 3, 5, 7 fm. Significant effects of the magnetic field are observed on the pion directed flow. The effects are larger at more forward and backward rapidities because of the larger velocity and hence larger Lorentz force. The effects increase with increasing impact parameter from 1 fm to 7 fm due to the increase of the magnetic field strength. The results are consistent with Fig.~\ref{y0-pxpy} (b):
The inner magnetic field changes the $\langle p_{x}\rangle$ of pions at large forward and backward rapidities. So $v_{1}$ of pions change accordingly. As seen from Fig.~\ref{y0-pxpy} (b), the effect of the magnetic field reduces the absolute values of $\langle p_{x}\rangle$ for both $\pi^{+}$ and $\pi^{-}$. Thus the $v_{2}$ of $\pi^{\pm}$ decreases at both forward and backward rapidities. This seems much more evident in Fig.~\ref{AuAub-pi-v1v2}, especially for $b=7$ fm.

\begin{figure}[htbp]
\centering
\includegraphics[angle=0,width=0.8\textwidth]{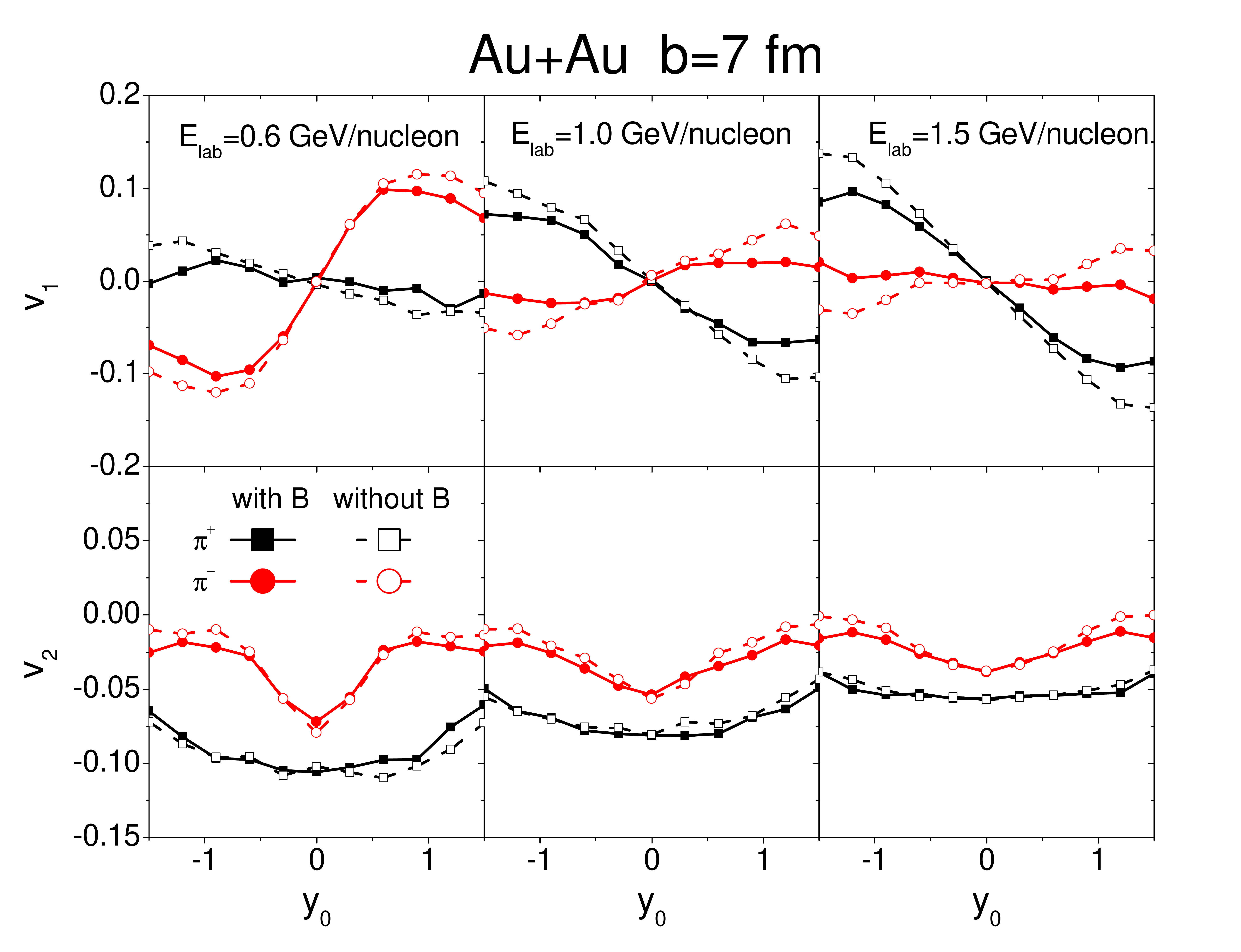}
\caption{\label{AuAuE-pi-v1v2} (Color online) $v_{1}$ and $v_{2}$ of charged pions as a function of normalized center-of-mass rapidity in Au+Au collisions at $E_{\text{lab}}$=0.6, 1, 1.5 GeV/nucleon with $b$=7 fm.
}
\end{figure}

\begin{figure}[htbp]
\centering
\includegraphics[angle=0,width=0.8\textwidth]{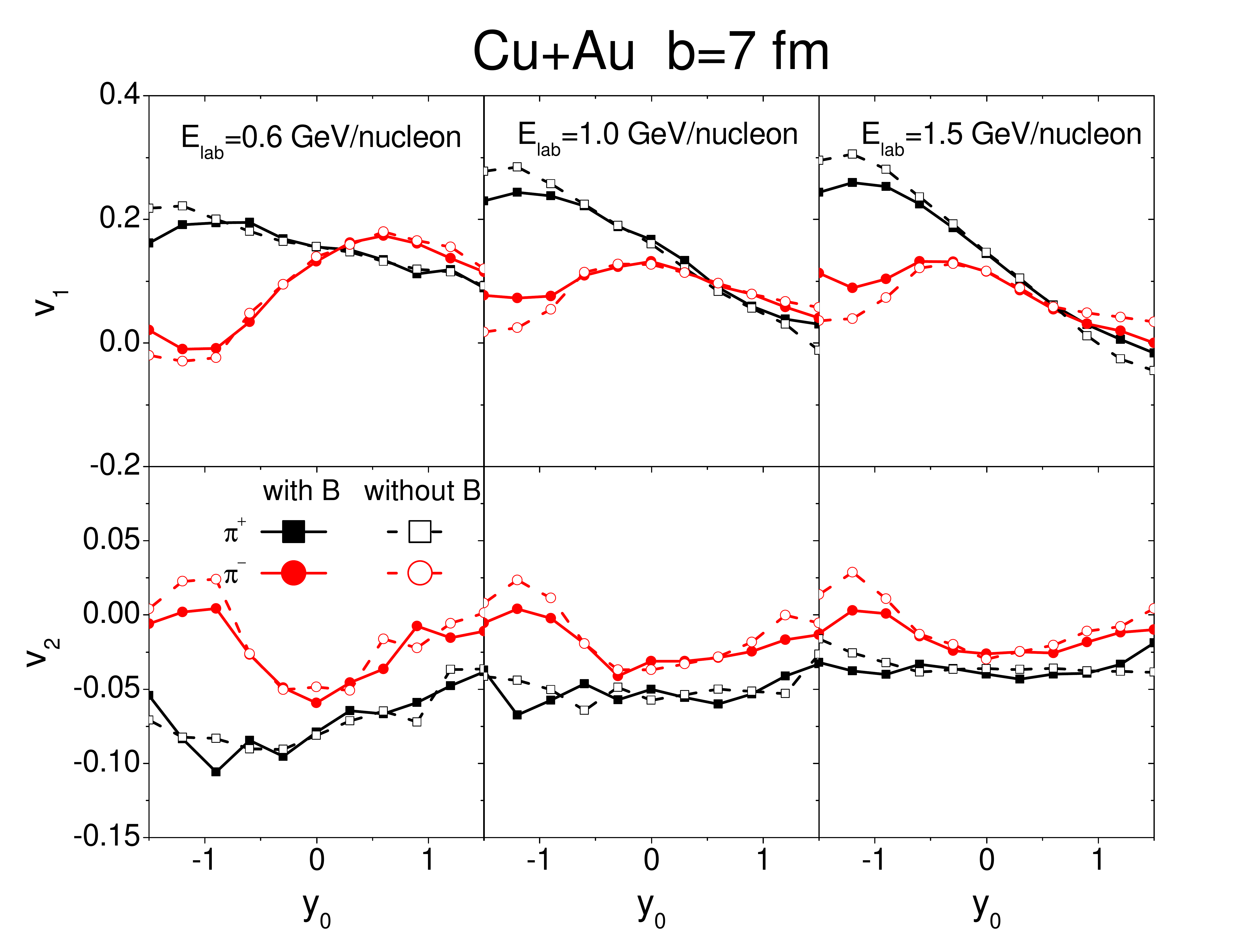}
\caption{\label{CuAuE-pi-v1v2} (Color online) $v_{1}$ and $v_{2}$ of charged pions as a function of normalized center-of-mass rapidity in Cu+Au collisions at $E_{\text{lab}}$=0.6, 1, 1.5 GeV/nucleon with $b$=7 fm.
}
\end{figure}

Figures~\ref{AuAuE-pi-v1v2} and~\ref{CuAuE-pi-v1v2} are $v_{1}$ and $v_{2}$ of charged pions as a function of normalized center-of-mass rapidity in Au+Au and Cu+Au collisions at $E_{\text{lab}}$=0.6, 1, 1.5 GeV/nucleon with $b$=7 fm. The effects of the magnetic field have a weak energy dependence. This is because of the interplay between the magnitude and the duration of the magnetic field, as shown in Figs.~\ref{fig6} and~\ref{E-intB}.

Our results of the magnetic field effect on particle anisotropic flows are qualitatively consistent with the IBUU model results of Ref.~\cite{Ou:2011fm} in that the flows change in the same direction. However, our UrQMD results are quantitatively different from those of IBUU because of the different physical ingredients in these two models, such as the equations of state and the two-bodies interaction cross sections. Nevertheless, It is clear from the calculations of both models that the pion directed flow is significantly affected by the magnetic field.

\section{Magnetic effects on isospin sensitive observables}

\begin{figure}[htbp]
\centering
\includegraphics[angle=0,width=0.8\textwidth]{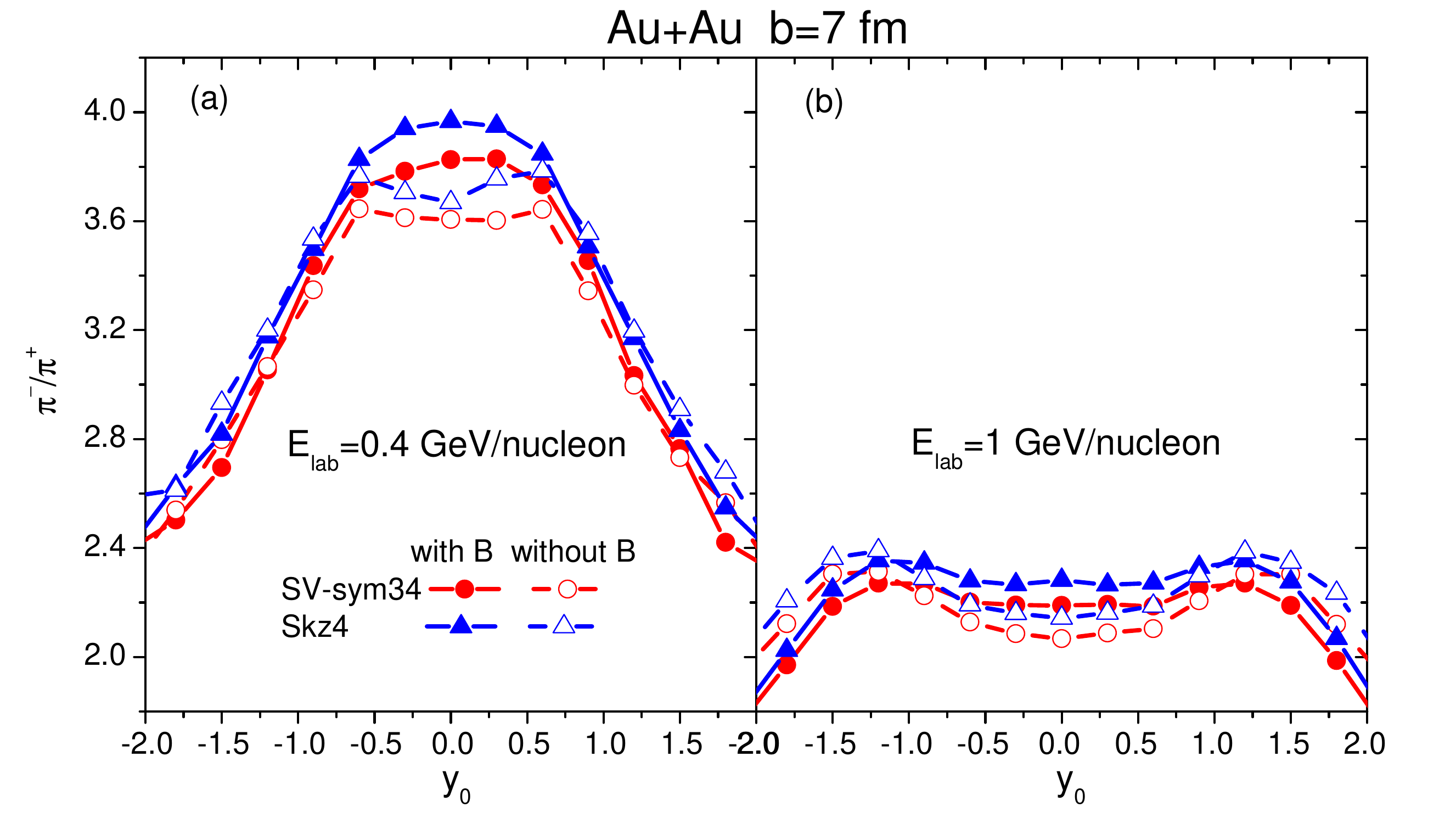}
\caption{\label{dn-ratio} (Color online) The $\pi^{-}/\pi^{+}$ yield ratio as a function of the normalized center-of-mass rapidity in Au+Au collisions at (a) $E_{\text{lab}}$=0.4 GeV$/$nucleon and (b) $E_{\text{lab}}$=1 GeV$/$nucleon with $b$=7 fm. Calculations with Skz4 (the corresponding slope parameter $L=5.7$ MeV) and SV-sym34 ($L=81$ MeV) interactions are shown.
}
\end{figure}

\begin{figure}[htbp]
\centering
\includegraphics[angle=0,width=0.8\textwidth]{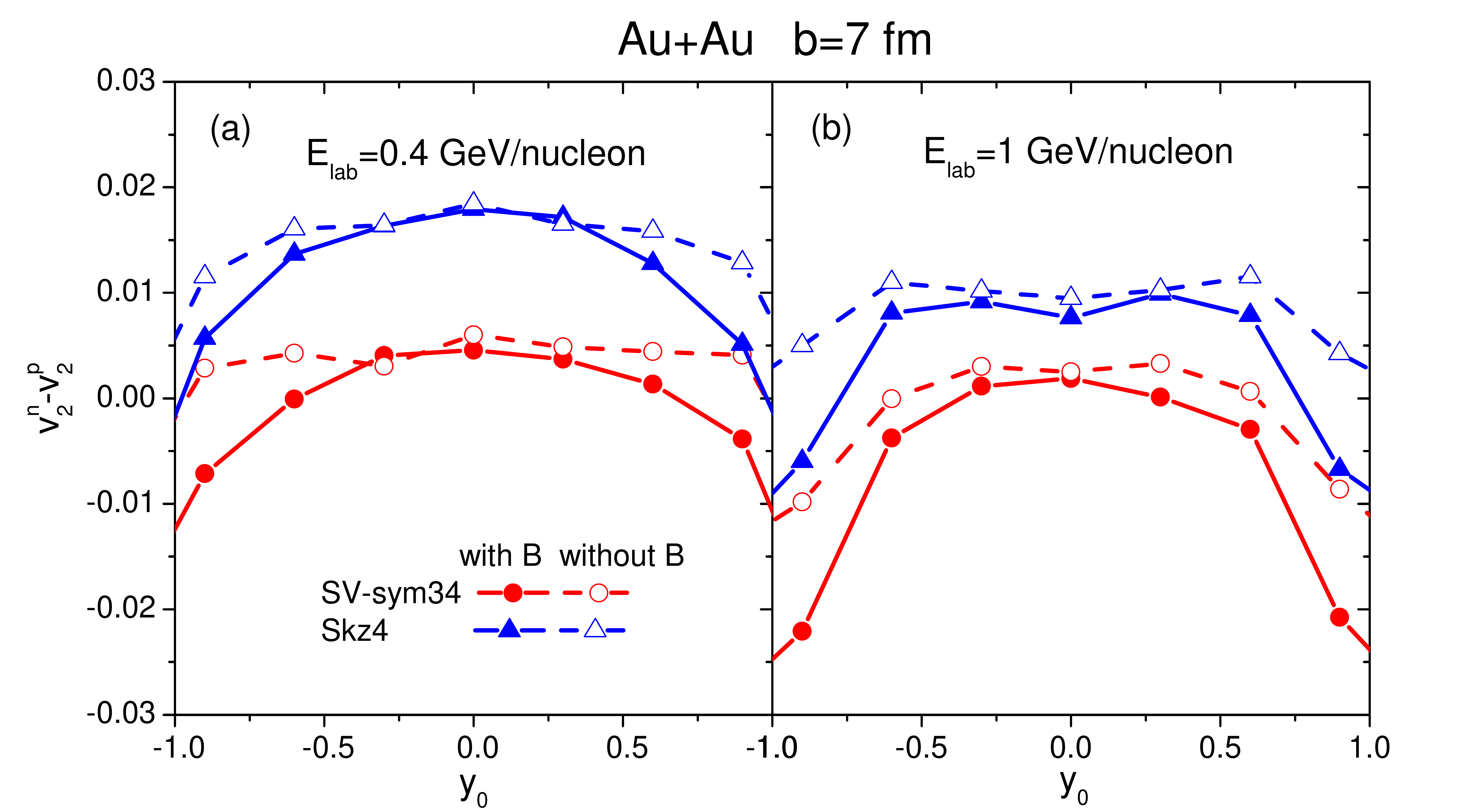}
\caption{\label{v2-diff} (Color online) The elliptic flow difference between neutrons and protons $v_{2}^{n}$-$v_{2}^{p}$ as a function of the normalized center-of-mass rapidity in Au+Au collisions at (a) $E_{\text{lab}}$=0.4 GeV/nucleon and (b) $E_{\text{lab}}$=1 GeV/nucleon with $b$=7 fm. Calculations with Skz4 ($L=5.7$ MeV) and SV-sym34 ($L=81$ MeV) interactions are shown.
}
\end{figure}

\begin{figure}[htbp]
\centering
\includegraphics[angle=0,width=0.9\textwidth]{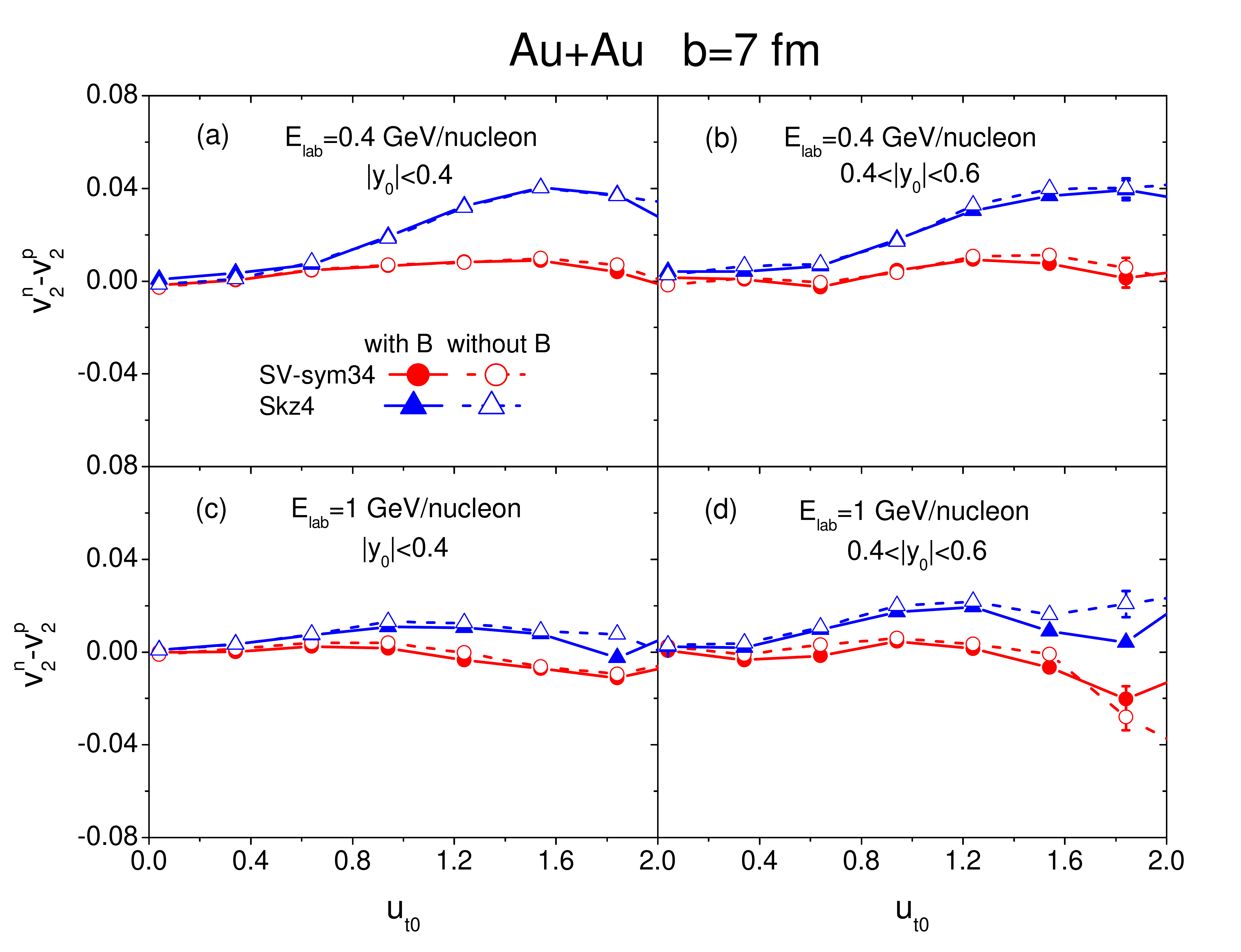}
\caption{\label{v2-diff-ut0} (Color online) The elliptic flow difference between neutrons and protons $v_{2}^{n}$-$v_{2}^{p}$ as a function of $u_{t0}$ in Au+Au collisions at (a,b) $E_{\text{lab}}$=0.4 GeV/nucleon and (c,d) $E_{\text{lab}}$=1 GeV/nucleon with $b$=7 fm. Calculations with Skz4 ($L=5.7$ MeV) and SV-sym34 ($L=81$ MeV) interactions are shown.
}
\end{figure}

It is well known that the $\pi^{-}/\pi^{+}$ yield ratio and the elliptic flow difference $v_{2}^{n}$-$v_{2}^{p}$ between neutrons and protons produced in HICs at intermediate energies are sensitive probes to the nuclear symmetry energy at high densities~\cite{Li:2002yda,Xiao:2008vm,Russotto:2011hq,Cozma:2013sja,yongjia2014prc}. We have shown in section IV that the motion of pions and protons are influenced by the magnetic field, thus it is interesting to examine the effect of magnetic field on the $\pi^{-}/\pi^{+}$ and $v_{2}^{n}$-$v_{2}^{p}$. Figs.~\ref{dn-ratio} and~\ref{v2-diff} shows the $\pi^-$/$\pi^+$ and $v_{2}^{n}$-$v_{2}^{p}$ as a function of rapidity calculated with different mean field potentials and with or without magnetic field. First, at both $E_{\text{lab}}$=0.4 and 1 GeV/nucleon, the magnetic field enhances the $\pi^-$/$\pi^+$ ratio at mid-rapidity region but depresses it at forward and backward rapidities, due to the magnetic focusing and defocusing effects on the positive and negative pions~\cite{Ou:2011fm}. We have checked that the total $\pi^-$/$\pi^+$ yield ratio remains the same for calculations with and without magnetic field, as it should and also consistent with the results reported in Ref.~\cite{Ou:2011fm}. Second, one sees clearly that the $\pi^-$/$\pi^+$ ratio in the mid-rapidity region obtained with Skz4 and without magnetic field is rather close to the result obtained with SV-sym34 and with magnetic field, illustrating that the effect of magnetic field on this observable is on the same order as the nuclear symmetry energy. This result is different from the result of Ref.~\cite{Ou:2011fm}, in which the symmetry energy effect on $\pi^-$/$\pi^+$ ratio is larger than the magnetic field effect. This difference may stem from different treatments of pion production in each model. As it became clear that, many physical effects (e.g., in-medium cross sections, pion dispersion relation, $\Delta$ production and decay, etc.) may significantly affect the sensitivity of $\pi^-$/$\pi^+$ ratio on the density-dependent symmetry energy~\cite{Cozma:2014yna,Li:2015hfa,Zhang:2017mps,Zhang:2017nck,Xu:2019hqg,Ono:2019ndq}, all these effects deserve further studies. For $v_{2}^{n}$-$v_{2}^{p}$, the effect of magnetic field is smaller than that of symmetry energy in the mid-rapidity region, especially at 0.4 GeV$/$nucleon, where one expect the effect of symmetry energy is more evident. While at forward and backward rapidities, the effect of the magnetic field is on the same order as the nuclear symmetry energy. $v_{2}^{n}$-$v_{2}^{p}$ calculated with magnetic field is smaller than that in absence of magnetic field, because magnetic field enhances the in-plane emission of positive charged particles, as observed in Fig. \ref{AuAu-p-v1v2}. In addition, the magnetic effects on the transverse-momentum-dependent flow difference $v_{2}^{n}$-$v_{2}^{p}$ are displayed in Fig.\ref{v2-diff-ut0}. The results from $E_{\text{lab}}$=0.4 and 1.0 GeV$/$nucleon within two rapidity windows are presented. Calculations with and without magnetic field track each other closely, it illustrates the weak effect of magnetic field on the transverse-momentum-dependent flow. Usually most of high-$u_{t0}$ particles are emitted very early and will escape the interaction zone very quickly in the collision, thus they will not strongly affected by the magnetic field due to short duration. Moreover, the $z$ component of the magnetic field is quite weak, thus the motion of charged particles in the transverse ($x$-$y$) plane can hardly be influenced. Our results suggest that comparing data without considering the magnetic field could lead to wrong conclusions. Effects of magnetic field should be included in future studies.

\section{Summary and Outlook}
In summary, within the transport model UrQMD, the time evolution and space distribution of internal magnetic field are calculated. The magnetic field strength is found to reach about $eB=470$ MeV$^{2}$ ($B=8\times10^{16}$ G) for Au+Au collision at $E_{\text{lab}}$=1 GeV/nucleon with impact parameter of 7 fm. The magnetic field in Cu+Au collisions exhibits somewhat different space distribution from that in Au+Au collisions. It is also found that the magnetic field has an effect on pion directed flow, significant at forward and backward rapidities, dependent of impact parameter. Our UrQMD calculation of the magnetic field effects on pion directed flow is qualitatively consistent with IBUU, but quantitatively different.
In addition, we found that the effects of the magnetic field on the $\pi^-$/$\pi^+$ ratio over the whole rapidity range and the elliptic flow difference $v_{2}^{n}$-$v_{2}^{p}$ between neutrons and protons at forward and backward rapidities are on the same order as those from the nuclear symmetry energy. While, the total $\pi^-$/$\pi^+$ yield ratio and transverse-momentum-dependent flow difference $v_{2}^{n}$-$v_{2}^{p}$ can hardly be affected by the magnetic field. In light of these results, the magnetic effects should be considered in future studies using these observables as a probe to the symmetry energy at super saturation densities.

\begin{acknowledgements}
We thank Dr. Haojie Xu for useful discussions. We acknowledge support by the computing server C3S2 in Huzhou University. The work is supported in part by the National Natural Science Foundation of China (Nos. 11405054, 11875125, 11847315, U1832139) and the U.S. Department of Energy (No. DE-SC0012910), and the Zhejiang Provincial Natural
Science Foundation of China (Grant No. LY18A050002 and LY19A050001).
\end{acknowledgements}

\bibliographystyle{unsrt}
\bibliography{ref}

\end{document}